%% file: main.tex
\documentclass[a4paper,11pt]{desyprocA4}

\usepackage[T1]{fontenc} % if needed
\usepackage{slashed}
\usepackage{booktabs}
\usepackage{feynmp}
\usepackage{amsmath}
\usepackage{placeins}
\usepackage{graphicx}
\usepackage{longtable}
\usepackage{epstopdf}
\renewcommand{\imath}{\mathrm{i}}

\usepackage{feynmp}
\usepackage[left=2cm,right=2cm,top=2cm,bottom=3cm]{geometry} 
 \usepackage[numbers,sort&compress]{natbib}
\usepackage[linktocpage=true]{hyperref}
\hypersetup{
  colorlinks   = true,    % Colours links instead of ugly boxes
  urlcolor     = blue,    % Colour for external hyperlinks
  linkcolor    = blue,    % Colour of internal links
  citecolor    = blue      % Colour of citations
}
\usepackage{tabu}
 
\begin{document} 
 
\title{\boldmath Two-loop correction to the Higgs boson mass in the MRSSM }

\author{Philip Diessner$^a$,  Jan Kalinowski$^b$, Wojciech Kotlarski$^{a,b}$\footnote{Corresponding author.} \ and Dominik St\"ockinger$^a$\\
[2ex]
$^a${Institut f\"ur Kern- und Teilchenphysik, TU Dresden, 
 01069 Dresden, Germany}\\
$^b${Faculty of Physics, University of Warsaw, Pasteura 5, 02093 Warsaw, Poland}}

\maketitle

\begin{abstract}
We present the impact of two-loop corrections on the mass of the lightest Higgs boson in the Minimal R-symmetric Supersymmetric Standard Model (MRSSM). 
These shift the Higgs boson mass up by typically
5~GeV or more. The dominant corrections arise from strong
interactions, from the gluon and its $N=2$ superpartners, the sgluon
and Dirac gluino, and these corrections further increase with large Dirac gluino mass. 
The two-loop contributions governed purely by Yukawa couplings and the MRSSM $\lambda,\Lambda$ parameters are smaller.
We also update an earlier analysis \cite{Diessner:2014ksa}, which
showed that the MRSSM can accommodate the measured Higgs and W boson
masses. Including the two-loop corrections increases the parameter space where the theory prediction agrees with the measurement. 

\end{abstract}

\section{Introduction}
\input{tex/introduction.tex}

\section{The MRSSM}
\input{tex/mrssm.tex}

\section{Higgs  mass dependence on the $\lambda,\Lambda$ superpotential parameters}
\input{tex/oldparameter.tex}

\section{QCD corrections and the two-loop corrected Higgs boson mass}
\input{tex/newparameter.tex}

\section{Update of benchmarks}
\input{tex/application.tex}

\input{tex/conclusions.tex}

\section*{Acknowledgments}
We would like to thank Kilian Nickel and Florian Staub for communication about \texttt{SARAH}. 
Work supported in part by the Polish National Science Centre
grants under OPUS-2012/05/B/ST2/03306, DEC-2012/05/B/ST2/02597,
  the European Commission through the contract PITN-GA-2012-316704 (HIGGSTOOLS), the German DFG Research
Training Group 1504 and the DFG grant STO 876/4-1.

\appendix

\bibliographystyle{ieeetr}
\bibliography{biblio}

\end{document}

%% file: tex/introduction.tex
The recent discovery at the LHC of a particle consistent with the long sought Higgs boson seemingly completes the Standard Model (SM). The mass of the particle is measured with an astonishingly high accuracy of $m_H=125.09\pm0.24$~GeV~\cite{Aad:2015zhl}.  
The precise determination of this mass is of paramount importance not only within the context of the Standard Model, but also for finding the path beyond it. In fact, a  number of experimental observations suggest that the SM cannot be the ultimate theory and many theoretical scenarios for the beyond the SM physics (BSM) have been proposed in past decades. 
In some models of BSM, in particular in supersymmetric extensions of the SM, the Higgs boson mass can be predicted. 
However, the current experimental accuracy is far better than theoretical predictions for Higgs boson mass in any given model of BSM physics.  
From the point of view of theory, the best accuracy has been achieved in the minimal supersymmetric extension of the SM (MSSM), in which the discovery of the Higgs boson and the determination of its mass have given a new impetus to the theoretical efforts. 
The most recent improvements comprise the inclusion of leading three-loop corrections \cite{Harlander:2008ju,Kant:2010tf}, resummations of leading logarithms beyond the two-loop level \cite{Hahn:2013ria,Draper:2013oza}, 
inclusion of the external momenta of two-loop self-energies \cite{Degrassi:2014pfa,Borowka:2014wla}, and the evaluation of the ${\cal O}(\alpha_t^2)$-contributions in the complex MSSM \cite{Hollik:2014wea,Hollik:2014bua}. The MSSM two-loop corrections controlled by Yukawa couplings and $\alpha_s$ have been known for quite some time for the real MSSM (see the above references for an overview of the literature).

The absence of any direct signal of supersymmetric particle production at the LHC, and the observed Higgs boson mass of $\sim$125 GeV being rather close to the upper value of $\sim$135 GeV achievable in the MSSM, are a strong motivation to consider non-minimal SUSY scenarios. 
In fact, non-minimal SUSY models can lift the Higgs boson mass (at the tree level by new $F$- or $D$-term contributions or at the loop level from additional new states), which makes these models more natural by reducing fine-tuning. 
They can also weaken SUSY limits either by predicting compressed spectra, or by reducing the expected missing transverse energy, or by reducing production cross-sections. The comparison of the measured Higgs boson mass  with the theoretically predicted values in any given model is therefore highly desirable. Although the theoretical calculations for the SM-like Higgs boson mass in such models are less advanced, progress is being made in the development of highly automated tools which greatly facilitate the computations in non-minimal SUSY models: \texttt{SARAH}~\cite{Staub:2010jh,Staub:2012pb,Staub:2013tta} automatically generates spectrum generators similar to \texttt{SPheno}~\cite{Porod:2003um,Porod:2011nf}; \texttt{FlexibleSUSY}~\cite{Athron:2014yba} automatically generates spectrum generators similar to \texttt{Softsusy}~\cite{Allanach:2001kg}.

In a recent paper \cite{Diessner:2014ksa} we considered the MRSSM, a highly motivated supersymmetric model with continuous R-symmetry \cite{Fayet:1974pd,Salam:1974xa} distinct from the MSSM. 
Since R-symmetry forbids soft Majorana gaugino masses as well as the higgsino mass term, additional superfields are needed. 
The MRSSM has been constructed in Ref.~\cite{Kribs:2007ac} as a minimal viable model of this type. It contains adjoint chiral superfields with R-charge 0 for each gauge sector and two additional Higgs weak iso-doublet superfields  with R-charge 2. 
Interestingly, R-symmetry also forbids large contributions to CP- and flavor-violating observables \cite{Buchmuller:1982ye,Kribs:2007ac}, so the MRSSM is generically in agreement with flavor data even for an anarchic flavor structure in the sfermion sector and for sfermion masses below the TeV scale.
Also, Dirac gluinos suppress the production cross-section for squarks, making squarks below the TeV scale generically compatible with LHC data.  Furthermore, models with R-symmetry and/or
Dirac gauginos contain promising dark matter
candidates \cite{Buckley:2013sca,Chun:2009zx,Belanger:2009wf}, and the collider
physics of the extra, non-MSSM-like states has been studied 
\cite{Fox:2002bu,Plehn:2008ae,Choi:2008ub,Choi:2010gc,Choi:2010an,Benakli:2012cy,ATLAS:2012ds,TheATLAScollaboration:2013jha,Kotlarski:2013lja}.

In Ref.~\cite{Diessner:2014ksa} the  complete next-to-leading order computation and discussion of the lightest Higgs boson and W boson masses has been performed (a similar analysis has been done in Ref.~\cite{Bertuzzo:2014bwa}). We showed that the model can accommodate measured values of these observables for interesting regions of parameter space with stop masses of order 1 TeV.  The outcome of the paper was not  obvious since in the MRSSM 
(i) the  lightest Higgs boson tree-level mass is typically reduced compared to the MSSM due to mixing with additional scalars, 
 (ii) the stop mixing is absent and (iii) R-symmetry necessarily introduces an SU(2) scalar triplet, which can increase $m_W$ already at the tree level. Nevertheless, we identified benchmark 
points BMP1, BMP2 and BMP3 illustrating different viable parameter regions for $\tan\beta=3,\, 10,\, 40$ respectively, and also verified that they are not excluded by further experimental constraints from Higgs observables, collider and low-energy physics. 

These promising results motivate a more precise computation of the Higgs boson mass in the MRSSM and a more precise parameter analysis.
Technically, this is facilitated by the Mathematica package \texttt{SARAH}, recently updated by providing \texttt{SPheno} routines, which calculate two-loop corrections to the CP-even Higgs scalars masses in the effective potential approximation and the gaugeless limit \cite{Goodsell:2014bna}. This is the level of precision of the established MSSM predictions except for the refinements mentioned above. It is also the level of precision at which the proof \cite{Hollik:2005nn} applies that the employed regularization by dimensional reduction preserves supersymmetry.
First applications of the improved \texttt{SARAH} version to the calculations 
of the Higgs boson masses in the R-parity violating MSSM \cite{Dreiner:2014lqa} and next-to-minimal SSM \cite{Goodsell:2014pla} have been published. 

Since in \cite{Diessner:2014ksa}
a judicious choice of the model parameters was needed to meet experimental constraints, and an estimate of unknown two-loop contributions was presented, it is of immediate interest to verify our findings at higher precision with the new \texttt{SARAH} version. 
The aim of the current paper is to calculate two-loop corrections for the Higgs boson mass in the MRSSM 
and present an update of the results obtained in~\cite{Diessner:2014ksa}.

The paper is organized as follows. 
 After a short  recapitulation of the MRSSM in section \ref{sec:mrssm}, we explain in section \ref{sec:old_parameters} our calculation framework and discuss the dependence of two-loop corrections on parameters that entered already at the one-loop level. 
The dependence on parameters that enter only at the two-loop level is investigated in section~\ref{sec:new_parameters}.  
In section~\ref{sec:application} we provide an update to the  analysis presented in \cite{Diessner:2014ksa} using the two-loop corrected masses of Higgses, before concluding in section~\ref{sec:conclusions}.

%% file: tex/mrssm.tex
\label{sec:mrssm}

The MRSSM has been constructed in Ref.~\cite{Kribs:2007ac} as a minimal supersymmetric model with unbroken continuous R-symmetry. The superpotential of the model reads as
\begin{align}
\nonumber W = & \mu_d\,\hat{R}_d \cdot \hat{H}_d\,+\mu_u\,\hat{R}_u\cdot\hat{H}_u\,+\Lambda_d\,\hat{R}_d\cdot \hat{T}\,\hat{H}_d\,+\Lambda_u\,\hat{R}_u\cdot\hat{T}\,\hat{H}_u\,\\ 
 & +\lambda_d\,\hat{S}\,\hat{R}_d\cdot\hat{H}_d\,+\lambda_u\,\hat{S}\,\hat{R}_u\cdot\hat{H}_u\,
 - Y_d \,\hat{d}\,\hat{q}\cdot\hat{H}_d\,- Y_e \,\hat{e}\,\hat{l}\cdot\hat{H}_d\, +Y_u\,\hat{u}\,\hat{q}\cdot\hat{H}_u\, ,
\label{eq:superpot}
 \end{align} 
where $\hat{H}_{u,d}$ are the MSSM-like Higgs weak iso-doublets, and $\hat{S},\, \hat{T},\, \hat{R}_{u,d}$ are  the singlet, weak iso-triplet and $\hat R$-Higgs weak iso-doublets, respectively. The usual MSSM $\mu$-term is forbidden; instead the $\mu_{u,d}$-terms involving R-Higgs fields are allowed. The $\Lambda,\lambda$-terms are similar to the usual Yukawa terms, where the $\hat R$-Higgs and $\hat{S}$ or $\hat{T}$ play the role of the quark/lepton doublets and singlets.

The usual soft mass terms of the MSSM scalar fields are allowed just like in the MSSM. In contrast, $A$-terms and  soft Majorana gaugino masses are  forbidden by R-symmetry. The fermionic components of the chiral adjoints, $\hat{\Phi}_i=\hat{\cal O},\,\hat{T}, \,\hat{S}$ 
for each standard model gauge group $i=SU(3)$, $SU(2)$, $U(1)$ respectively, are paired with standard gauginos $\tilde{g},\tilde{W},\tilde{B}$ 
to build Dirac fermions and the corresponding mass terms. The Dirac gaugino masses generated by D-type spurions produce  additional terms with the auxiliary $\mathcal{D}$-fields in the Lagrangian,
\begin{align}
V_D =& M_B^D (\tilde{B}\,\tilde{S}-\sqrt{2} \mathcal{D}_B\, S)+
M_W^D(\tilde{W}^a\tilde{T}^a-\sqrt{2}\mathcal{D}_W^a T^a)+
M_g^D(\tilde{g}^a\tilde{O}^a-\sqrt{2}\mathcal{D}_g^a O^a)
+ \mbox{h.c.}\,,
\label{eq:potdirac}
\end{align}
which after being eliminated through their equations of motion, lead to the appearance of Dirac masses in the scalar sector as well. 
The soft-breaking scalar mass terms read
\begin{align}
V^{EW}_{SB}= \, &  m_{H_d}^2 (|H_d^0|^2 + |H_d^-|^2) +m_{H_u}^2 (|H_u^0|^2 + |H_u^+|^2)+ [B_{\mu}(H_d^- H_u^+- H_d^0 H_u^0 ) + \mbox{h.c.} ] \nonumber
\\%[2mm] 
\ &
+m_{R_d}^2 (|R_d^0|^2 + |R_d^+|^2)   +m_{R_u}^2 |R_u^0|^2+m_{R_u}^2 |R_d^-|^2 
\nonumber \\ &
+m_S^2 |S|^2 +m_T^2 |T^0|^2 +m_T^2 |T^-|^2 +m_T^2 |T^+|^2 +m_O^2 |O|^2\nonumber 
\\
  &+\tilde{d}^*_{L,{i }}  m_{q,{i j}}^{2} \tilde{d}_{L,{j }} 
 +\tilde{d}^*_{R,{i }}  m_{d,{i j}}^{2} \tilde{d}_{R,{j }}  +\tilde{u}^*_{L,{i }} m_{q,{i j}}^{2} \tilde{u}_{L,{j }} +\tilde{u}^*_{R,{i }}  m_{u,{i j}}^{2} \tilde{u}_{R,{j }}
 \nonumber \\ 
 & +\tilde{e}^*_{L,{i}} m_{l,{i j}}^{2} \tilde{e}_{L,{j}} +\tilde{e}^*_{R,{i}} m_{e,{i j}}^{2} \tilde{e}_{R,{j}} +\tilde{\nu}^*_{L,{i}} m_{l,{i j}}^{2} \tilde{\nu}_{L,{j}} \,.
\label{eq:othersoftpot}
\end{align}
The electroweak symmetry breaking (EWSB)
is triggered by non-zero vacuum expectation values of the $R=0$ neutral EW scalars, which are parameterized as
\begin{align*} 
H_d^0=& \, \textstyle{\frac{1}{\sqrt{2}}} (v_d + \phi_{d}+i  \sigma_{d}) \;,& 
H_u^0=& \, \textstyle{\frac{1}{\sqrt{2}}} (v_u + \phi_{u}+i  \sigma_{u}) \;,\\ 
T^0  =& \, \textstyle{\frac{1}{\sqrt{2}}} (v_T + \phi_T +i  \sigma_T)   \;,&
S   = & \, \textstyle{\frac{1}{\sqrt{2}}} (v_S + \phi_S +i  \sigma_S)   \;;
\end{align*} 
R-Higgs bosons carry R-charge 2 and therefore do not develop vacuum expectation values. We stress that in general the mixing of $\phi_T,\phi_S$ with $\phi_u$ and $\phi_d$ leads to a reduction of the lightest Higgs boson mass at the tree-level compared to the MSSM.

%% file: tex/oldparameter.tex
\label{sec:old_parameters}
\begin{figure}[t]
\centering
\begin{minipage}{0.3\textwidth}
\includegraphics[width=\textwidth]{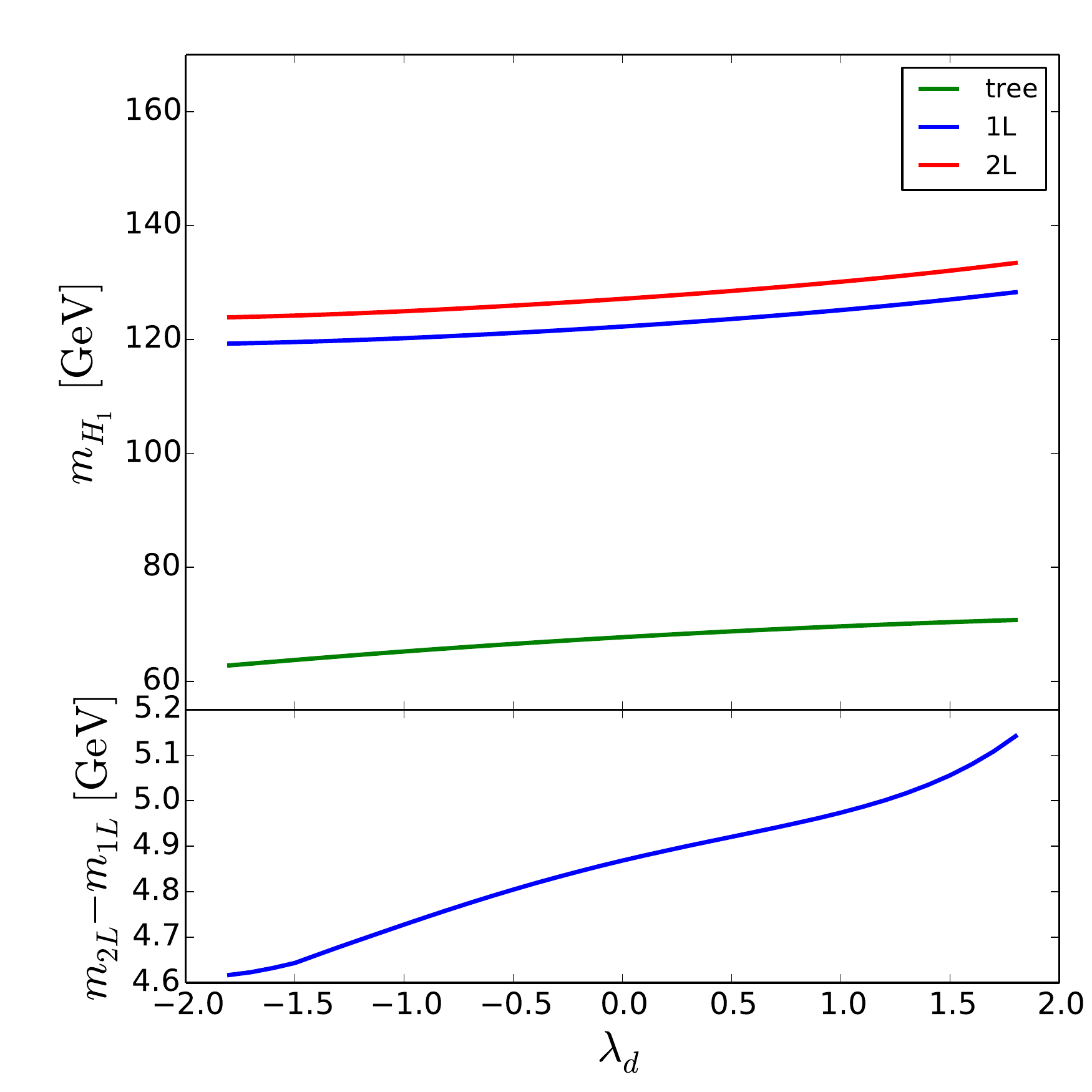}
\end{minipage}
\begin{minipage}{0.3\textwidth}
\includegraphics[width=\textwidth]{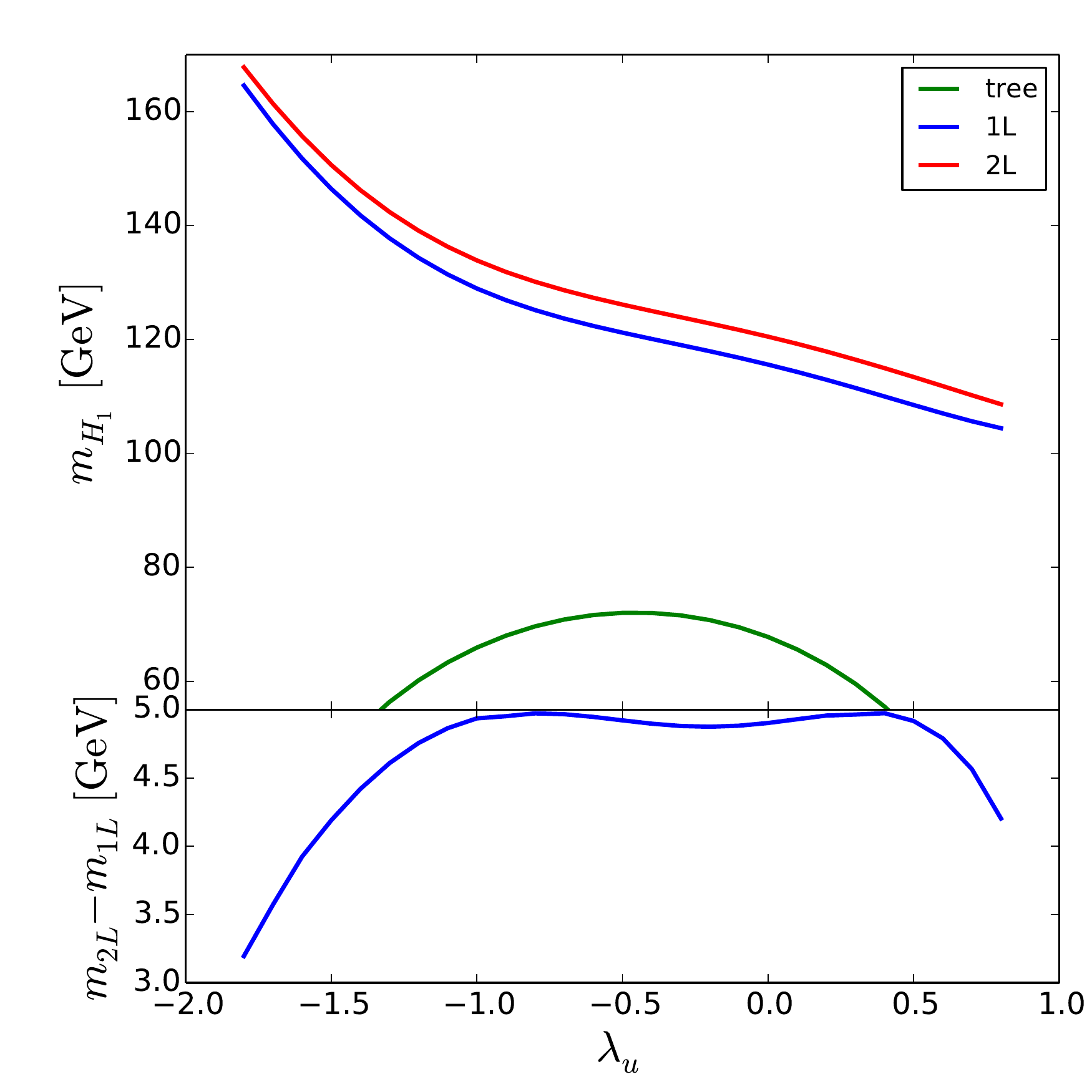}
\end{minipage}
\begin{minipage}{0.3\textwidth}
\includegraphics[width=\textwidth]{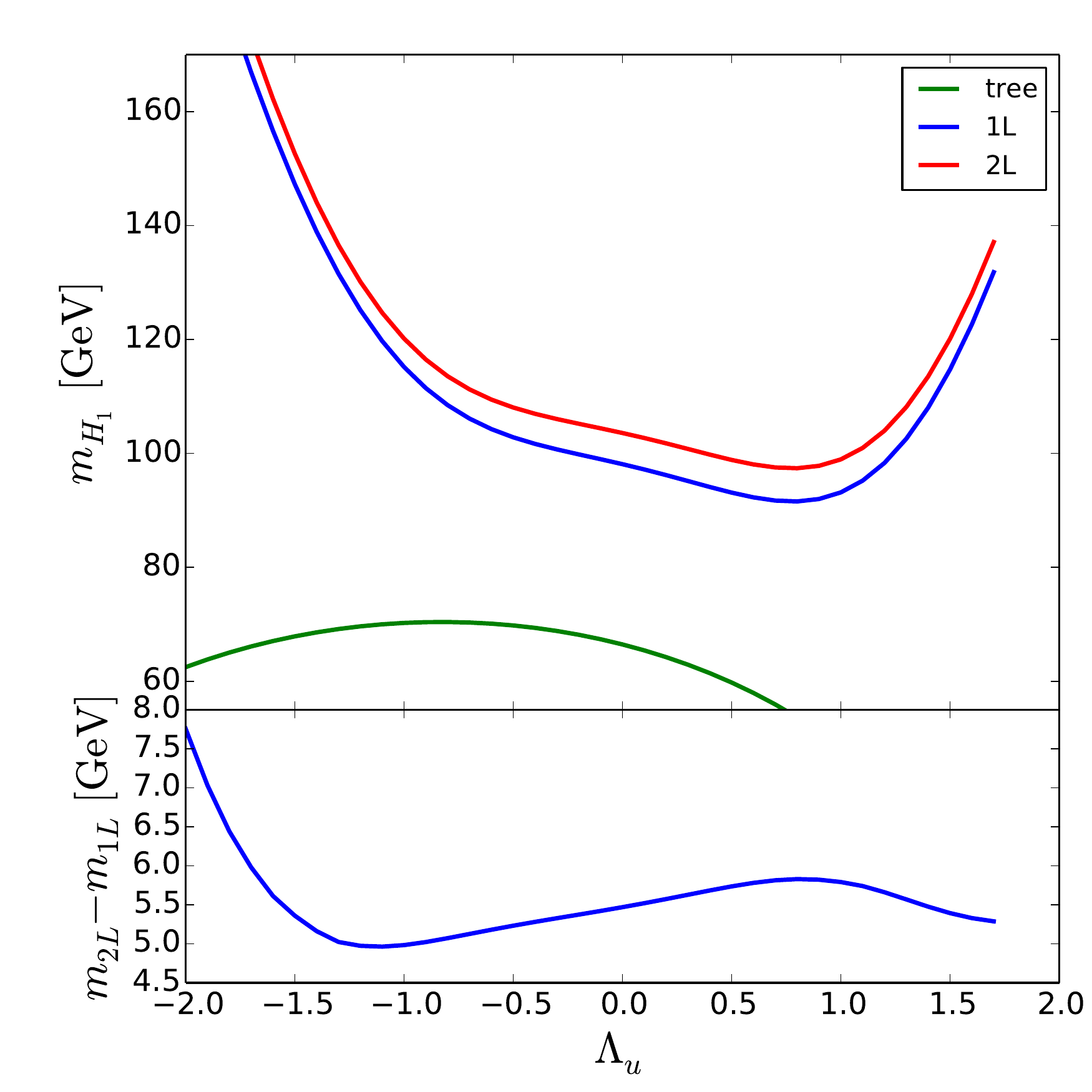}
\end{minipage}
\caption{Lightest MRSSM Higgs boson mass $m_{H_1}$, and the difference $m_{2L}-m_{1L}$ between masses calculated at the two-loop and one-loop level, as a function of $\lambda_d$, $\lambda_u$, 
$\Lambda_u$, respectively.  In the upper parts of the figure lines from top to bottom correspond to two-loop, one-loop and tree level calculations. 
All other parameters are set to the values of benchmark point BMP1 with $\tan\beta=3$ (see Tab.~\ref{tab:BMPold}).}
\label{img:dhlamdlamuTest1}
\end{figure}

\begin{figure}[t]
\centering
\begin{minipage}{0.3\textwidth}
\includegraphics[width=\textwidth]{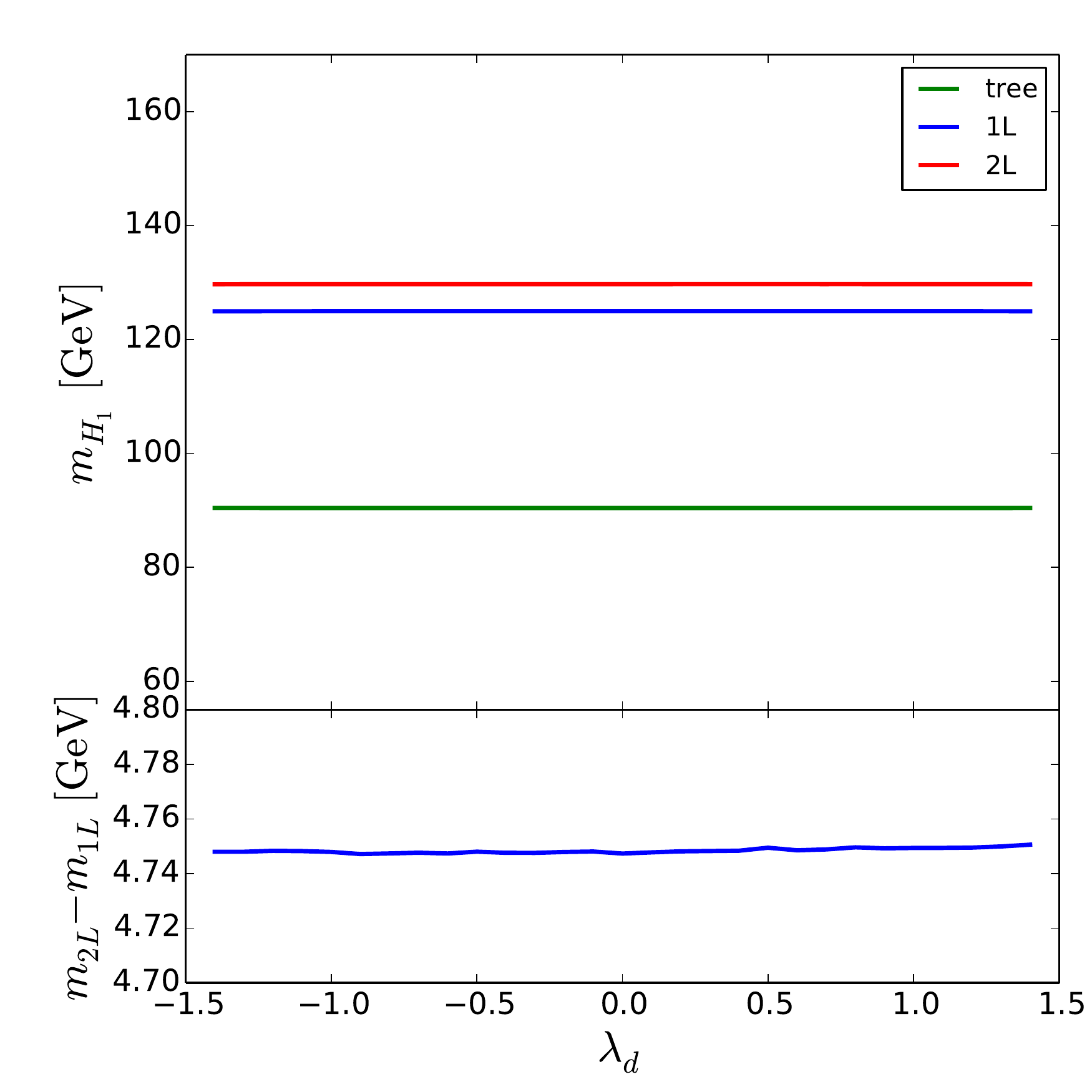}
\end{minipage}
\begin{minipage}{0.3\textwidth}
\includegraphics[width=\textwidth]{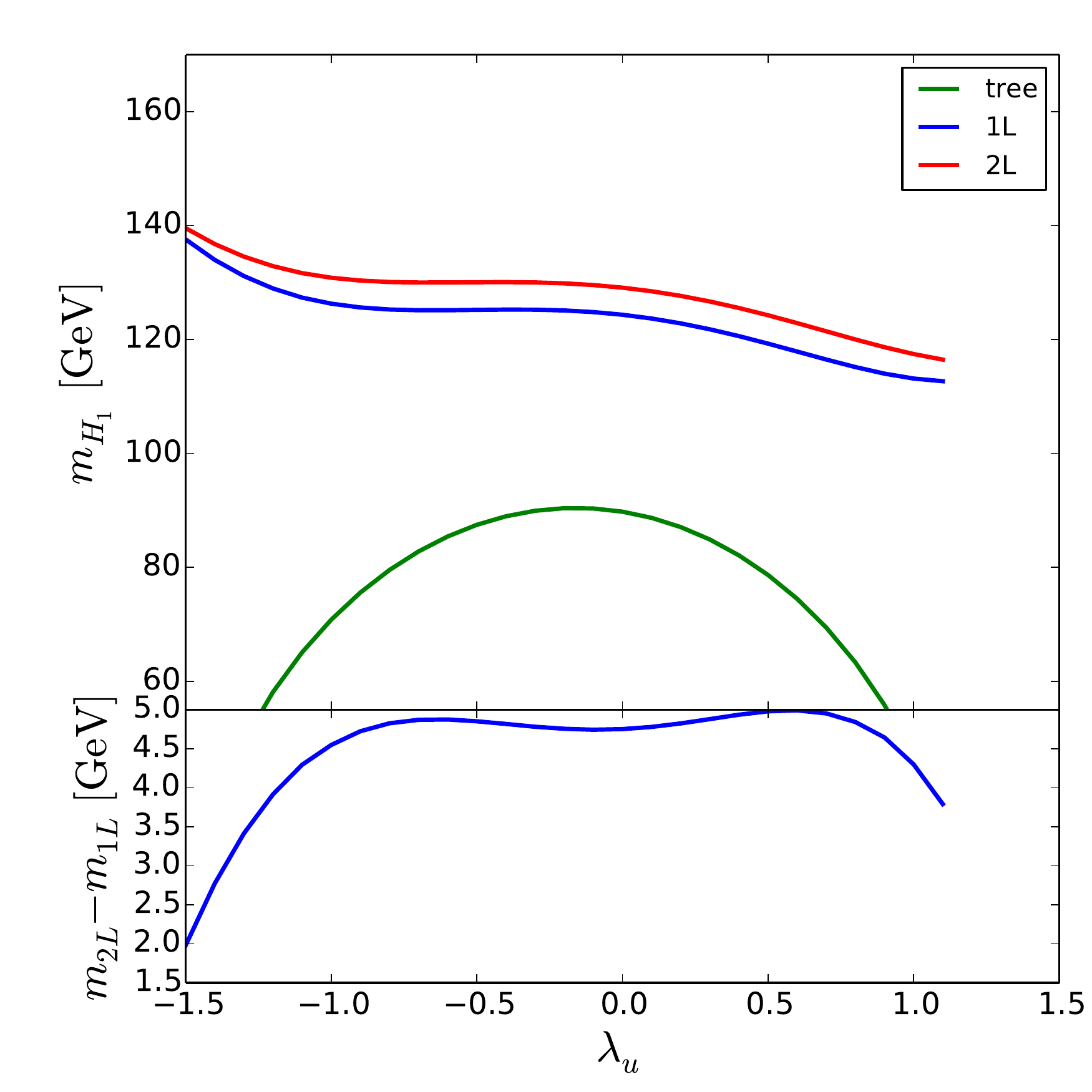}
\end{minipage}
\begin{minipage}{0.3\textwidth}
\includegraphics[width=\textwidth]{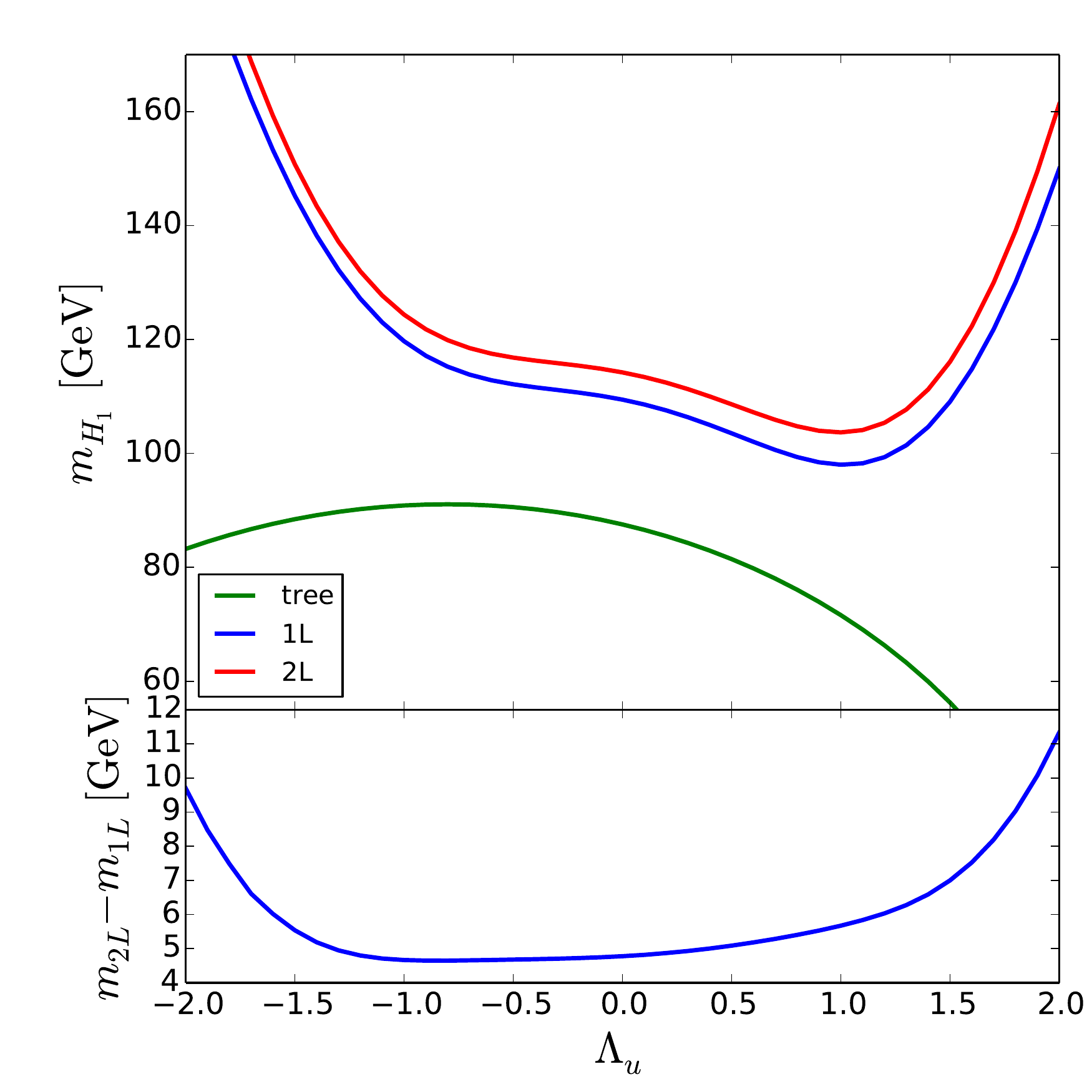}
\end{minipage}
\caption{As in Fig.~\ref{img:dhlamdlamuTest1}, but for benchmark point BMP3 with $\tan\beta=40$ (see Tab.~\ref{tab:BMPold}).}
\label{img:dhlamdlamuTest2}
\end{figure}

We now present the MRSSM Higgs boson mass prediction at the two-loop level. We use the same renormalization scheme as in Ref.~\cite{Diessner:2014ksa}, where all SUSY parameters are defined in the $\overline{\text{DR}}$ scheme and $m_{H_d}^2$, $m_{H_u}^2$, $v_S$ and $v_T$ are determined by  minimizing the effective potential at the two-loop order. 
The discussion is divided into two parts. In the present section we begin with the one-loop contributions, which are dominated by terms of ${\cal O}(\alpha_{t,b,\lambda})$, where $\alpha_\lambda$ collectively denotes squares of the superpotential couplings $\lambda_{u,d}$ and $\Lambda_{u,d}$. We then discuss the two-loop contributions of ${\cal O}(\alpha_{t,b,\lambda}^2)$, i.e.\ ones which depend on parameters which already play a role at the one-loop level. In the subsequent section we then discuss those two-loop corrections which involve new parameters.

In the usual MSSM, the one-loop contributions to the Higgs boson mass are dominated by top/stop contributions. In the MRSSM, these contributions are also important, but they are simpler since stop mixing is forbidden by R-symmetry (corresponding to the MSSM parameter $X_t \equiv A_t-\mu/\tan\beta=0$). This implies that the top/stop contributions cannot reach values as high as in the MSSM for a given stop mass scale. However, as mentioned above, the MRSSM superpotential contains new terms governed by $\lambda_{u,d}$ and $\Lambda_{u,d}$ which have a Yukawa-like structure. References~\cite{Diessner:2014ksa,Bertuzzo:2014bwa} have given a useful analytical approximation for these contributions.
 In the the limit $\lambda=\lambda_u=-\lambda_d$, $\Lambda=\Lambda_u=\Lambda_d$, $v_S\approx v_T\approx0$ and large $\tan\beta$,  we get
\begin{align}
\label{eq:effpothiggs}
&\Delta m_{H_1,\text{eff.pot},\lambda}^2 =\frac{2 v^2}{16 \pi^2}\left[\frac{\Lambda^2\lambda^2}{2}   +  \frac{4\lambda^4 + 4\lambda^2 \Lambda^2 + 5 \Lambda^4}{8} \log \frac{m^2_{R_u}}{Q^2} \right. \nonumber \\
  &\phantom{MM}+\left(\frac{\lambda^4}{2}-\frac{\lambda^2\Lambda^2}{2} \frac{m^2_S}{m^2_T-m^2_S}\right)\log \frac{m^2_S}{Q^2}
% \\ &
+\left(\frac{5}{8}\Lambda^4+\frac{\lambda^2\Lambda^2}{2} \frac{m^2_T}{m^2_T-m^2_S}\right)\log\frac{m^2_T}{Q^2}
\\& \phantom{MM}- \left.\left(\frac{5}{4}\Lambda^4-\lambda^2\Lambda^2 \frac{(M^D_W)^2}{(M^D_B)^2-(M^D_W)^2}\right)\log \frac{(M^D_W)^2}{Q^2}
%\\&
 - \left(\lambda^4+\lambda^2\Lambda^2 \frac{(M^D_B)^2}{(M^D_B)^2-(M^D_W)^2} 
 \right )\log\frac{(M^D_B)^2}{Q^2} \right]\;.\nonumber
\end{align} 
This result shows a behavior proportional to $\lambda^4$, $\Lambda^4$ and $\log m_{\text{soft}}^2$. This 
is similar  to
the top/stop contributions  as the $\lambda$'s and $Y_t$ appear in a similar fashion in superpotential.

We expect therefore that the two-loop result will depend on these model parameters (which already entered at the one-loop level) 
in a manner similar  to the pure top quark/squarks two-loop contributions, i.e.\ similar to the MSSM ${\cal O}(\alpha_t^2)$ contributions without stop mixing. 

In Figs.~\ref{img:dhlamdlamuTest1} and \ref{img:dhlamdlamuTest2} the dependence of the lightest Higgs boson mass 
calculated at tree-, one- and  two-loop levels for two benchmarks BMP1 and BMP3 on different model parameters 
is shown.  All parameters except the ones shown on the horizontal axes are set to the values of the benchmark points
defined in Ref.~\cite{Diessner:2014ksa} (see Tab.~\ref{tab:BMPold}). Indeed the $\lambda,\Lambda$ behavior of the two-loop corrections is very similar to the one of the corresponding one-loop corrections.  The numerical impact of the two-loop 
$\lambda$, $\Lambda$-contributions is rather small, typically less than $1~$GeV, except for very large 
$|\lambda_u|,|\Lambda_u|>1$, where they can reach several GeV. 
Particularly, the strong $\lambda_u$ dependence for large $\lambda_u$ is already manifest for the tree-level mass; 
this is due to the mixing with the singlet state already present in the tree-level mass matrix.

One should remember that very large one-loop contributions are required to bring the predicted Higgs boson mass close to the experimental one. In the preferred parameter regions, the $\lambda$, $\Lambda$ are large but still moderate enough not to blow up the two-loop contributions.

Overall, the total two-loop contributions (including the ones to be
discussed in the subsequent section) are in the range between $4$ and $5$~GeV, except in the very large $\lambda,\Lambda$ regions. This is in agreement with the estimate given in Ref.~\cite{Diessner:2014ksa}, and it confirms the validity of the perturbative expansion in spite of the large one-loop corrections.

\FloatBarrier

%% file: tex/newparameter.tex
\label{sec:new_parameters}

\begin{figure}[tb!]
\centering
\begin{tabu} to 0.9\linewidth {
  X[1,c,m]
  X[1.2,l]    % controls horizontal spacing
  X[1,c,m]
  X[1,l] } 
\begin{fmffile}{img/feynman_rules/FeynDia31} 
\fmfframe(20,20)(20,20){ 
\begin{fmfgraph*}(75,75) 
\fmfleft{l1}
\fmfright{r1,r2}
\fmf{dashes}{l1,v1}
\fmf{scalar}{r1,v1}
\fmf{scalar}{v1,r2}
\fmflabel{$O_{S,{\alpha}}$}{l1}
\fmflabel{$\tilde{t}_{{i \beta}}$}{r1}
\fmflabel{$\tilde{t}^*_{{i \gamma}}$}{r2}
\end{fmfgraph*}} 
\end{fmffile}
 & \hspace*{-8mm}
$i g_3 M^{D}_O \lambda^{\alpha}_{\gamma\beta} (\delta_{iR} - \delta_{iL} )\,, \quad$
&
 \begin{fmffile}{img/feynman_rules/FeynDia310} 
 \fmfframe(20,20)(20,20){ 
 \begin{fmfgraph*}(75,75) 
 \fmfleft{l1}
 \fmfright{r1,r2}
 \fmf{fermion}{v1,l1}
 \fmf{fermion}{r1,v1}
 \fmf{scalar}{r2,v1}
 \fmflabel{$\bar{t}_{i,{\alpha}}$}{l1}
 \fmflabel{${\tilde{g}}_{D,{\beta}},{\tilde{g}}_{D,{\beta}}^C$}{r1}
 \fmflabel{$\tilde{t}_{i,{\gamma}}$}{r2}
 \end{fmfgraph*}} 
 \end{fmffile} 
& \hspace*{-8mm}
$ \frac{i}{\sqrt{2}} g_3 \lambda^{\beta}_{\alpha\gamma}(\delta_{iR} - \delta_{iL} )\, .$

 \end{tabu}
\caption{Feynman rules needed to evaluate diagrams of Fig.~\ref{fig:feynman_diagrams}. In the right diagram, the charge-conjugated gluino ${\tilde{g}}_{D,{\beta}}^C$ applies in the case of $i=L$, ${\tilde{g}}_{D,{\beta}}$ in the case of $i=R$.}
 \label{fig:feynman_rules}
\end{figure}
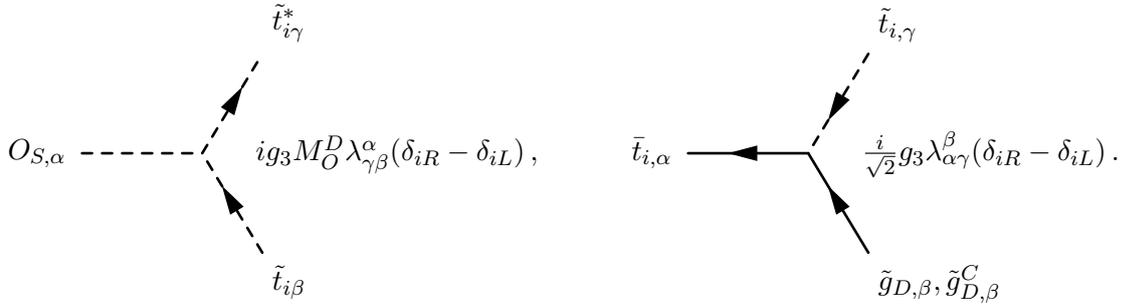

At two-loop level the strongly interacting sector and the strong coupling $\alpha_s$ appear directly in the Higgs boson mass predictions. These two-loop corrections involve not only the gluon but also the Dirac gluino and the sgluon, the scalar component of the octet superfield $\hat{O}$. They can be expected to be sizable, and they depend on the 
gluino Dirac mass and sgluon soft mass parameters.\footnote{
These parameters already play a role at lower order, appearing in corrections to $Y_t$ (through threshold corrections to $\hat{\alpha}_s$), though the influence on, for example, $\overline{\text{DR}}$ top mass is negligible.}
The gluino Dirac mass parameter $M_O^D$ appears not only directly as the gluino mass but, via Eq.~(\ref{eq:potdirac}), also in couplings and mass terms of sgluons, inducing the mass splitting\footnote{In Ref.~\cite{Diessner:2014ksa} a simplifying assumption was made that masses of the scalar and pseudoscalar components of (complex) sgluon field were equal, since it was unimportant for that analysis. }   between the real and imaginary parts of the sgluon 
field, $O=\textstyle{\frac{1}{\sqrt 2}} (O_S+i O_A)$.  The  masses of the scalar sgluons $O_S$ and pseudoscalar  
 sgluons $O_A$ are related by the tree-level formula $m^2_{O_S} = 4 (M_O^D)^2 + m^2_{O_A}$, where $m_{O_A}^2$ is equal to the soft-breaking 
parameter $m^2_{O}$ \cite{Fox:2002bu,Choi:2008ub}. The relevant
vertices and Feynman rules are depicted in
Fig.~\ref{fig:feynman_rules}. We assume real $M_O^D$, so only the
scalar $O_S$ acquires the direct coupling to sfermions proportional to
$M_O^D$, via
Eq.~(\ref{eq:potdirac}).

The  structure of the strong corrections is thus markedly different from the MSSM case, where only the Majorana gluino and the gluon appear. In the following, we study the magnitude and the behavior of the corrections as a function of the parameters $M_O^D$ and $m_O^2$.

\subsection{Analytic formulas}

As in the previous section, we begin with an analytic approximation for the leading contributions of 
${\cal O}(\alpha_t\alpha_s)$, i.e.\ two-loop strong corrections proportional to $Y_t^2$. This provides 
us with qualitative insight and serves as a check of the code.
Generally, in the gaugeless limit the two-loop corrections from gluinos and sgluons contribute only to 
the diagonal part of the $\{\phi_d, \phi_u\}$ submatrix of the scalar Higgs boson mass matrix. 
In the MRSSM the ${\cal O}(\alpha_t\alpha_s)$ terms  contribute only to the $\phi_u \phi_u$ element. This already 
constitutes a difference to the MSSM, where the $\mu$-term violates R-symmetry and Peccei-Quinn symmetry 
leading to couplings of stops to $\phi_d$.

\begin{figure}[tb]
  \begin{center}
    \includegraphics[width=0.24\textwidth]{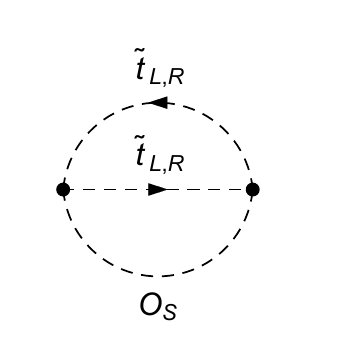}
    \includegraphics[width=0.48\textwidth]{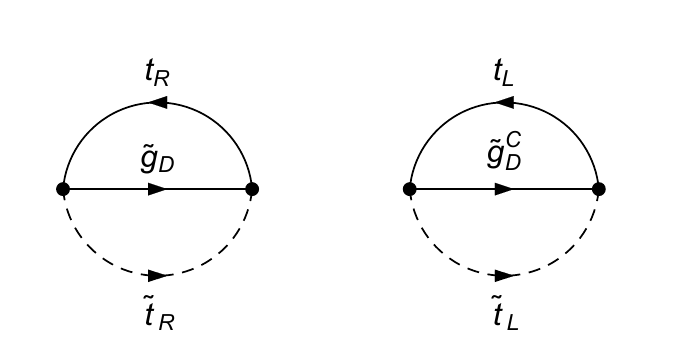}
  \end{center}
  \caption{Two-loop diagrams contributing  to the Higgs boson mass via
    Eq.~(\ref{eq:1}) that depend on the Dirac mass $M_O^D$ and the
    soft sgluon mass $m_O$. We only draw diagrams involving top/stop;
    similar diagrams exist for all quark/squark flavors. }
  \label{fig:feynman_diagrams}
\end{figure}

Figure~\ref{fig:feynman_diagrams} shows two-loop diagrams contributing to the Higgs boson mass at ${\cal O}(\alpha_t\alpha_s)$ 
that explicitly depend
 on $m_O$ and/or $M_O^D$. 
These diagrams provide the following contribution to the effective potential:
\begin{eqnarray}
 V^{(2)}_{eff} &=& \frac{8 g_3^2}{(16\pi^2)^2} (M_O^{D})^2 \sum_{i=L,R} f_{SSS} (m^2_{\tilde{t}_i}, m^2_{\tilde{t}_i}, m^2_{O_S} ) + \frac{8 g_3^2}{(16\pi^2)^2} \sum_{i=L,R} f_{FFS} (m_t^2, m_{\tilde t_i}^2, m_{\tilde g_D}^2 )\, ,
 \label{eq:1}
\end{eqnarray}
where the functions $f_{SSS}$ and $f_{FFS}$ are defined in \cite{Martin:2001vx}.  
The effective potential $V^{(2)}_{eff}$ depends on $v_u$ through stop masses, which in the gaugeless limit approach
\begin{eqnarray}
  m_{\tilde t_L \tilde t_L}^2 &\to& m_q^2 + \frac{1}{2} Y_t^2 v_u^2  \, ,  \\
  m_{\tilde t_R \tilde t_R}^2 &\to& m_u^2 + \frac{1}{2} Y_t^2 v_u^2 \, .
\end{eqnarray}
Equation~(\ref{eq:1}) can be obtained from Ref.~\cite{Martin:2001vx} by applying translation rules from real fields to complex ones. Many such rules can be found in Ref.~\cite{Goodsell:2014bna}; an additional rule needed here for the case of a Lagrangian $\mathcal L \ni -c \Phi_1 |\Phi_2|^2$, where $\Phi_1,c \in \mathbb{R}, \Phi_2 \in \mathbb{C}$, is $
  V_{SSS} = \textstyle{\frac{1}{2}} |c|^2 f_{SSS} (m_1^2, m_2^2, m_2^2)$.

An important difference to the MSSM is that contributions with fermion mass insertions, corresponding to $\overline{FF}S$-type contributions in Ref.~\cite{Martin:2001vx}, are not present in the MRSSM. Such contributions vanish due to the lack of L-R mixing between squarks.
Hence the gluino mass appears in a simpler way than in the MSSM. Likewise, the sgluon only enters via the $SSS$-type diagram of Fig.~\ref{fig:feynman_diagrams}. An $SS$-type diagram  vanishes due to the color structure. 

The corresponding two-loop contribution to the $\phi_u \phi_u$ Higgs boson mass matrix element in zero-momentum approximation is then given by\footnote{As pointed out in \cite{Goodsell:2014bna}, in \texttt{SARAH} and \texttt{SPheno} the two-loop tadpole contributions are included directly in vacuum minimization condition and not in Eq.~\ref{eq:2}.}
\begin{equation}
  \left [ \Delta m_{H_1}^2 \right ]_{\phi_u \phi_u} = \left ( \frac{\partial^2 }{\partial v_u \partial v_u} - \frac{1}{v_u} \frac{\partial }{\partial v_u} \right ) V^{(2)}_{eff}\, .
  \label{eq:2}
\end{equation}
For large $\tan \beta$, corrections of order $\mathcal O (\alpha_b \alpha_s)$ cannot be neglected any more. 
But since they contribute only to $\phi_d \phi_d $ matrix element, their impact  on mass of the lightest Higgs, which stems mainly from the $\phi_u \phi_u$ element, is small.
Results of Eq.~(\ref{eq:1}) where compared with the results  of two-loop routines from the \texttt{SARAH}-generated \texttt{SPheno} module.

%----------------------------------------------------------------
\subsection{Numerical analysis}
\begin{figure}[th]
\centering
\begin{minipage}[t]{0.31\textwidth}
\includegraphics[width=\textwidth]{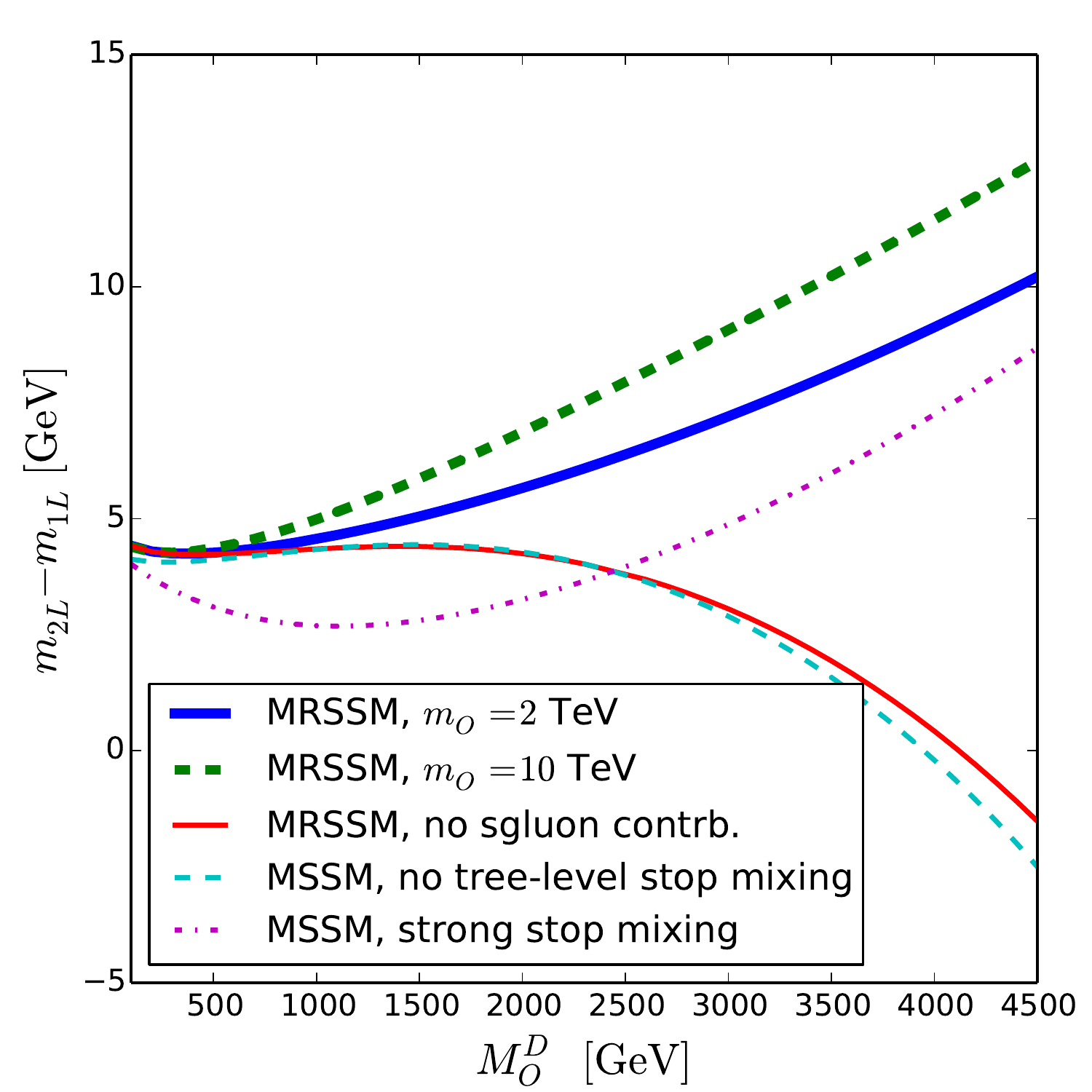}
\end{minipage}
\begin{minipage}[t]{0.31\textwidth}
\includegraphics[width=\textwidth]{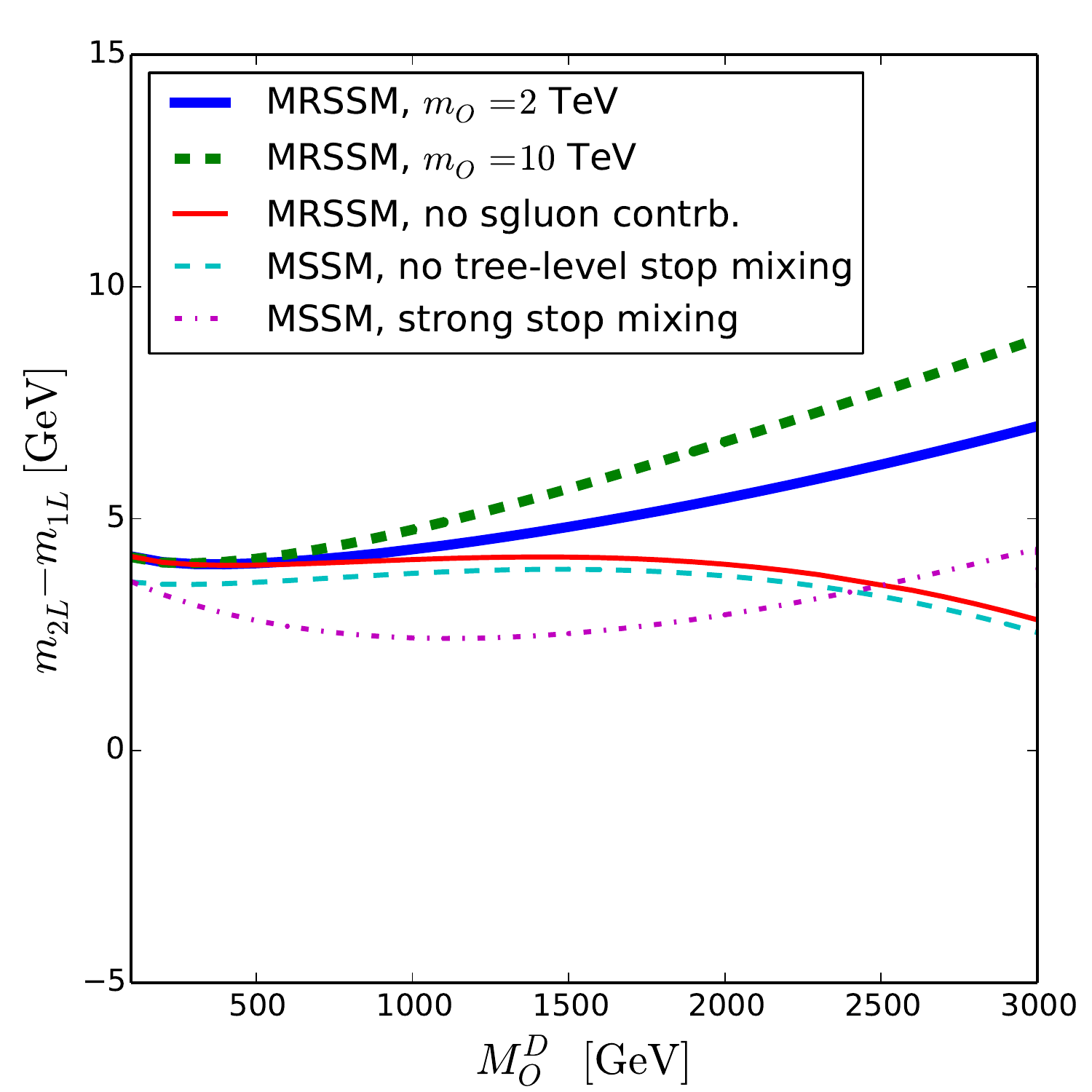}
\end{minipage}
\begin{minipage}[t]{0.31\textwidth}
\includegraphics[width=\textwidth]{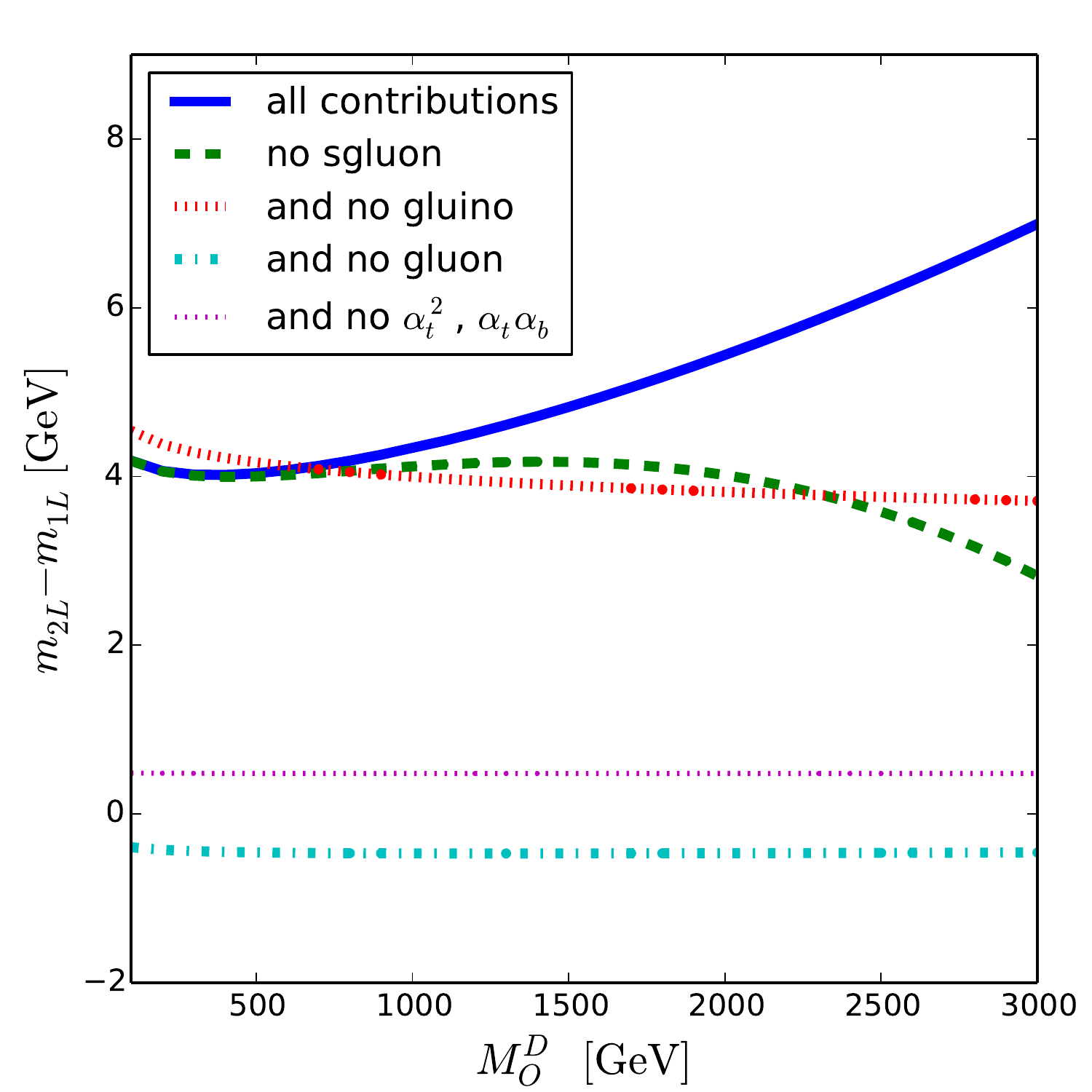}
\end{minipage}
\caption{Two-loop contributions to the SM-like Higgs boson mass depending on the gluino mass in the 
MRSSM for BMP1 (left panel) and BMP3 (middle panel) and two different values of the soft sgluon mass parameter 
$m_O=2$~TeV (thick solid blue line) and $10$~TeV (thick dashed green line) with all contributions, respectively, and without
the sgluon contributions (thin solid red line). 
For comparison also the MSSM contributions for no (thin dashed light
blue line) and maximal (purple dotted line) stop mixing are plotted. The chosen MSSM parameters are given in Tab.~\ref{tab:MSSM}. 
For BMP3 and $m_O=2$~TeV the right panel shows the result, when successively switching off
dominating and sub-dominating contributions.
}
\label{img:mgmocontrb}
\end{figure}

\begin{table}[tb!]
\begin{center}
\begin{tabular}{l|rrrrrrrrrr}
            & $\tan\beta$ & $M_1$ & $M_2$ & $\mu$ & $m_A$ & $m^2_{q,u,d;(3,3)}$ & $m^2_{q,u,d}$ & $m^2_{l,e}$  & $A_{\tau,b}$ & $X_t$\\
\midrule
left panel  & 3           & 600   & 500  & 400   & 700   & $1000^2$              & $2000^2$        & $1000^2$    & 0          & 0/2000\\
right panel & 40          & 250   & 500  & 400   & 700   & $1000^2$              & $2000^2$        & $1000^2$    & 0          & 0/2000\\
\end{tabular}
\end{center}
\caption{Definition of the fixed parameters for the MSSM points in Fig. \ref{img:mgmocontrb}. All parameters in GeV or GeV${}^2$, where appropriate. The
stop mixing parameter $X_t$ is given both for the cases of no and large stop mixing.}\label{tab:MSSM}
\end{table}
We now turn to the numerical analysis of the complete two-loop corrections to
the SM-like Higgs boson mass, using the full evaluation within the framework of \texttt{SARAH} and \texttt{SPheno}. 
The first two panels of Fig.~\ref{img:mgmocontrb} focus on the gluino
and sgluon mass dependence, which arises mainly from the ${\cal
O}(\alpha_{t}\alpha_s)$ corrections; they show the two-loop corrections as a function of the gluino mass parameter for two different values of
the soft sgluon mass, $m_O = 2$ and 10 TeV for two benchmarks BMP1 and BMP3; other parameters are fixed at benchmark values.   
For comparison, the two-loop result without the sgluon contribution is shown as well 
(i.e.\ without the first diagram of
Fig.~\ref{fig:feynman_diagrams}). We also plot the MSSM prediction
with strong stop mixing and without any sfermion mixing the at tree-level. 

The first two panels show that the dependence in the MRSSM without sgluon
contributions is very similar to the one in the MSSM without stop
mixing. The corresponding thin solid red and thin dashed light blue curves in
Fig.~\ref{img:mgmocontrb} show a characteristic drop for large gluino
masses.
This is understandable as in the MSSM without sfermion mixing the
gluino contribution is precisely the same as in the MRSSM and given by
the two corresponding diagrams in Fig.~\ref{fig:feynman_diagrams}. 
The Dirac or Majorana nature of the gluino does not matter since the Dirac partner, 
the octet superfield $\hat{O}$ has no direct couplings to quark superfields. 
A few TeV gluino masses slightly increase the Higgs boson mass, but for
larger values of $M_O^D$ the $f_{FFS}$ function becomes
negative and drives the correction downwards.

In the full MRSSM calculations, including the sgluon diagrams 
strongly changes the behavior.  Surprisingly, the full MRSSM two-loop
contributions resemble the MSSM contributions with large stop
mixing. In both cases, large gluino masses strongly enhance the Higgs
boson mass, however for different reasons. In the MSSM the increase can be traced back to the
additional  $\overline{FF}S$-type diagram which is directly
proportional to $M_O^D$ and which vanishes in the limit of no
stop-mixing.  In the MRSSM, on the other hand, the sgluon diagram
grows with $M_O^D$ both due to the sgluon-stop-stop coupling, which
scales like $M_O^D$, 
 and to an increase in the scalar (but not pseudoscalar) sgluon
 mass.
Due to the sgluon contributions the total two-loop contributions
to the Higgs boson mass in the MRSSM are larger than the ones in the
MSSM. They are further increased by heavy sgluons.

The third panel of Figure~\ref{img:mgmocontrb} compares the 
numerical impact of individual contributions by successively switching
off contributions. It allows to read off the contributions from
sgluon, gluino and gluon, of ${\cal
O}(\alpha_{t}^2,\alpha_t\alpha_b)$, and the
remaining two-loop contributions (particularly the $\lambda,\Lambda$
contributions). The gluon diagrams alone contribute 
approximately $+4~$GeV. The 
negative gluino and the positive sgluon corrections together amount to
an additional upward shift of the Higgs boson mass, which can reach
several GeV for large Dirac gluino masses. The remaining contributions are far
smaller and amount to around $-1~$GeV for the ${\cal O}(\alpha_{t}^2,\alpha_t\alpha_b)$
contributions and $+0.5$~GeV for the remaining contributions.

\FloatBarrier

%% file: tex/application.tex
\label{sec:application}
\begin{figure}[th]
\centering
\begin{minipage}{0.3\textwidth}
\includegraphics[width=\textwidth]{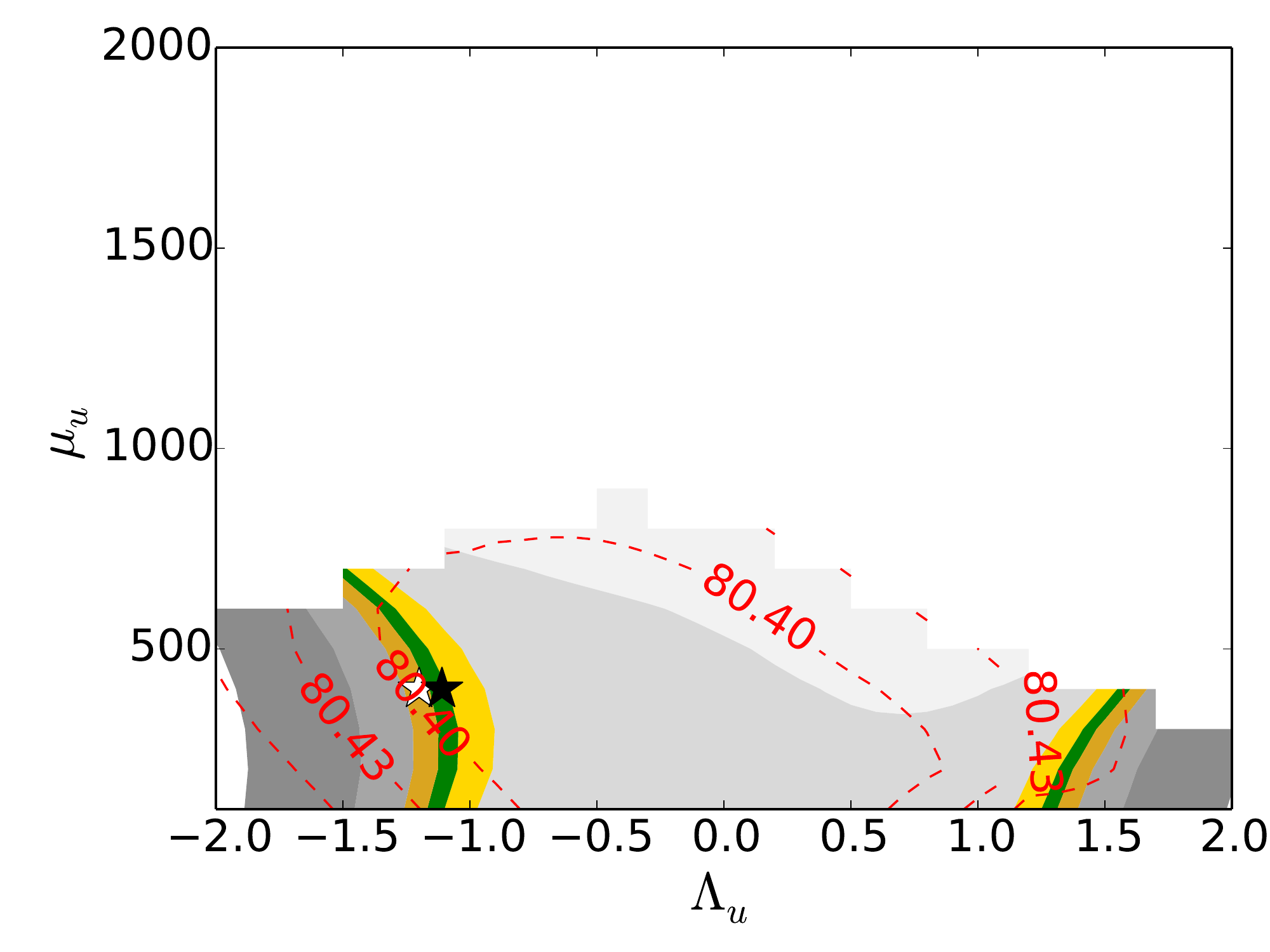}\\
\includegraphics[width=\textwidth]{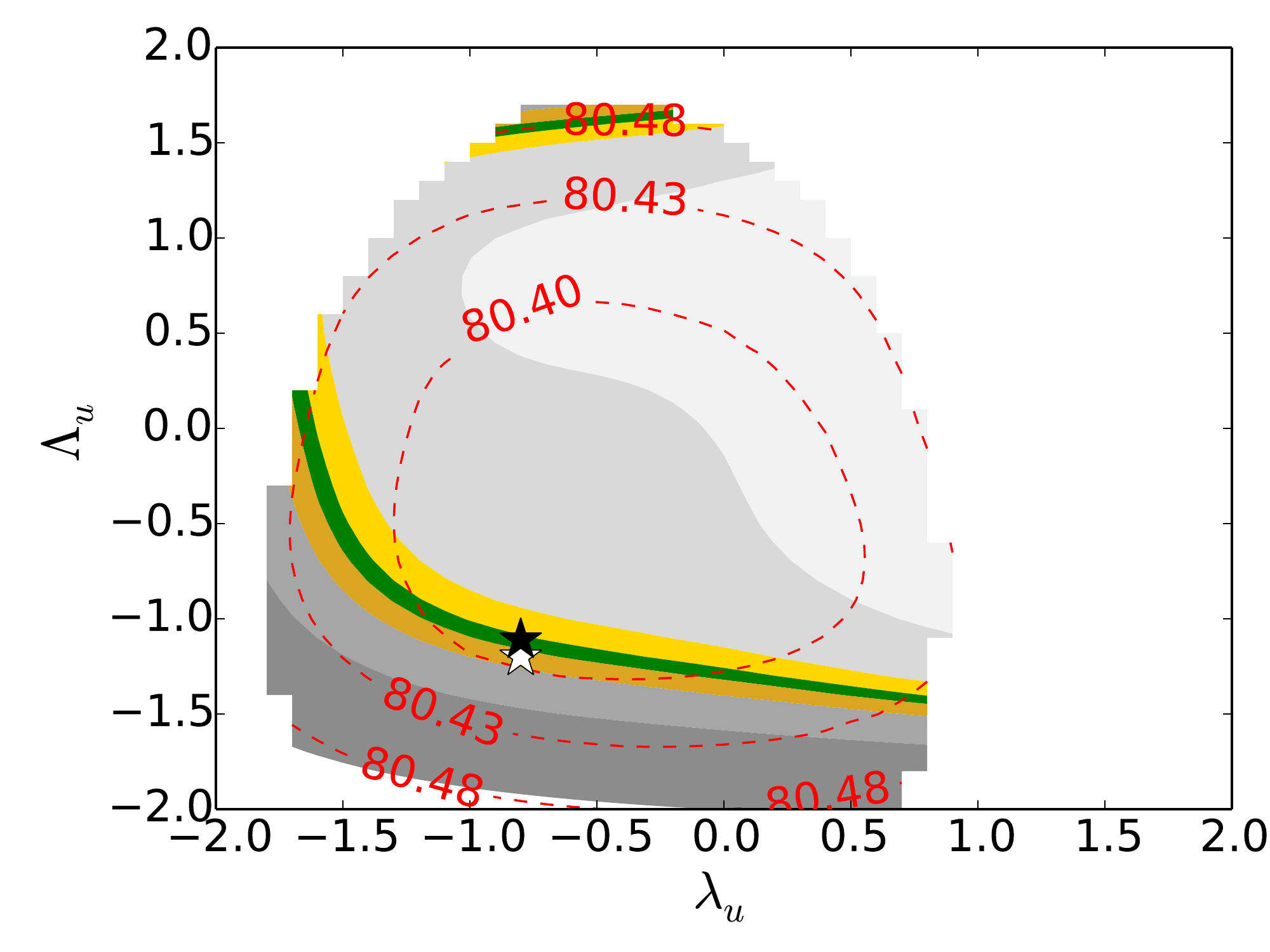}
\end{minipage}
\begin{minipage}{0.3\textwidth}
\includegraphics[width=\textwidth]{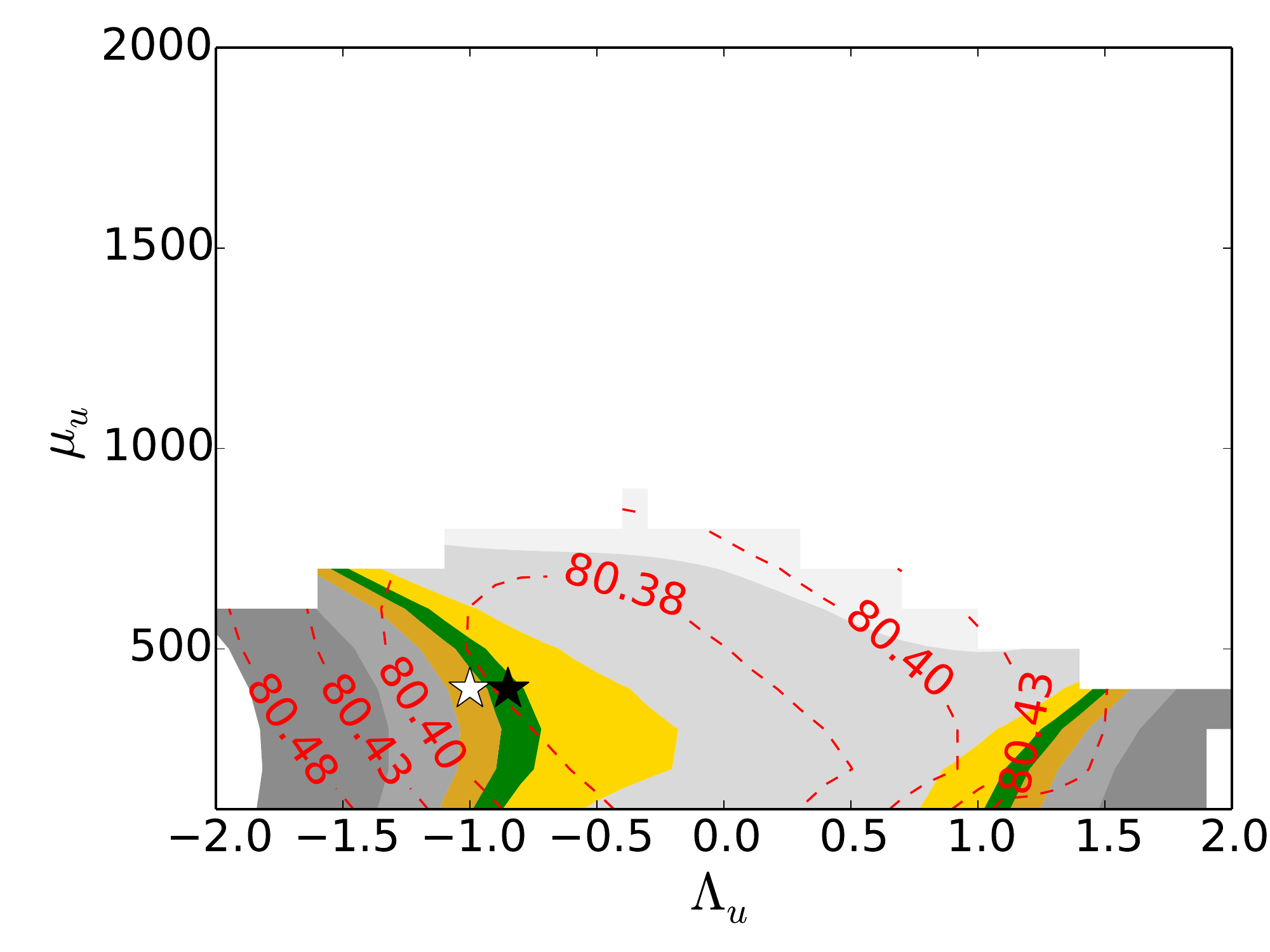}\\
\includegraphics[width=\textwidth]{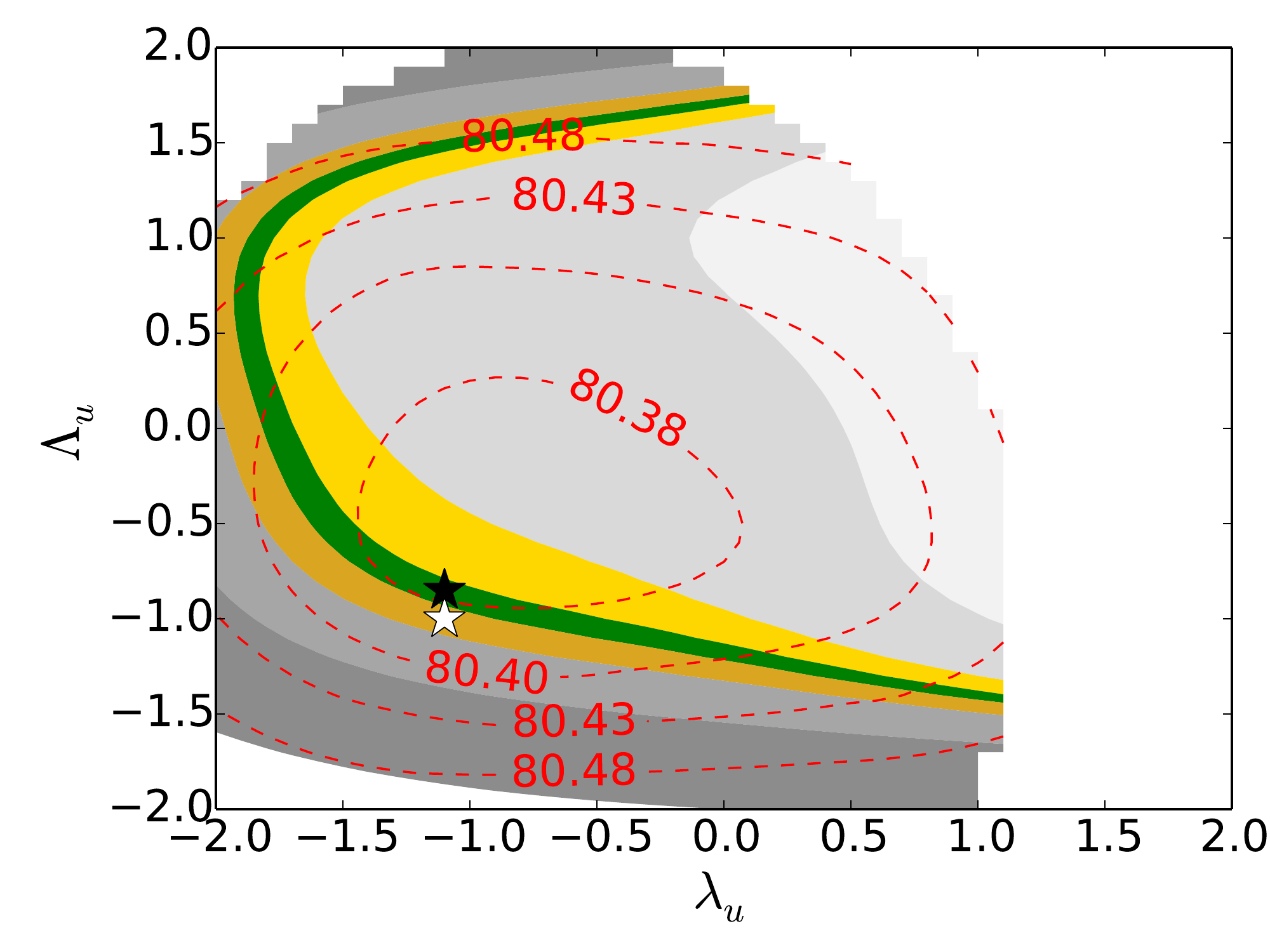}
\end{minipage}
\begin{minipage}{0.3\textwidth}
\includegraphics[width=\textwidth]{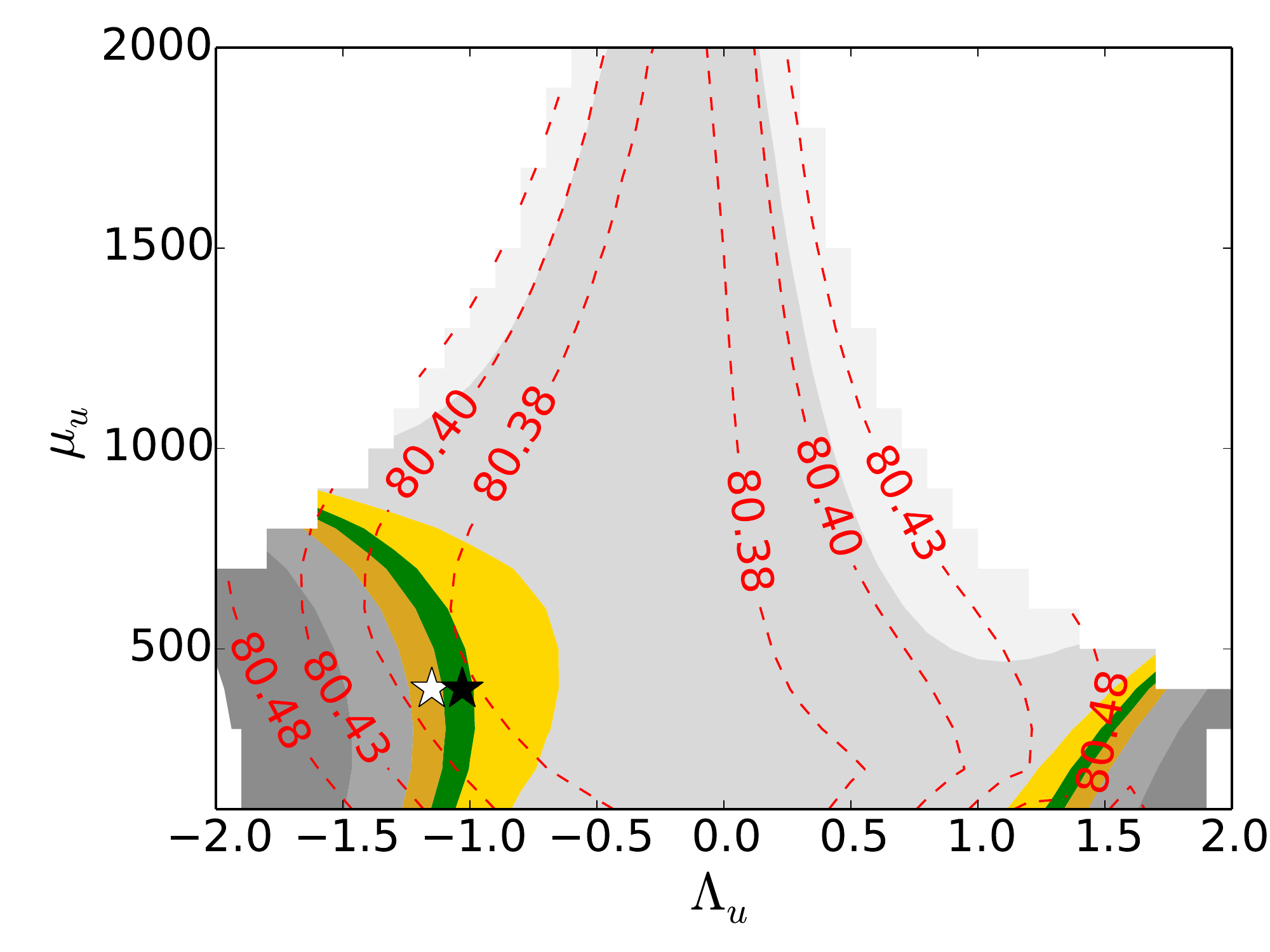}\\
\includegraphics[width=\textwidth]{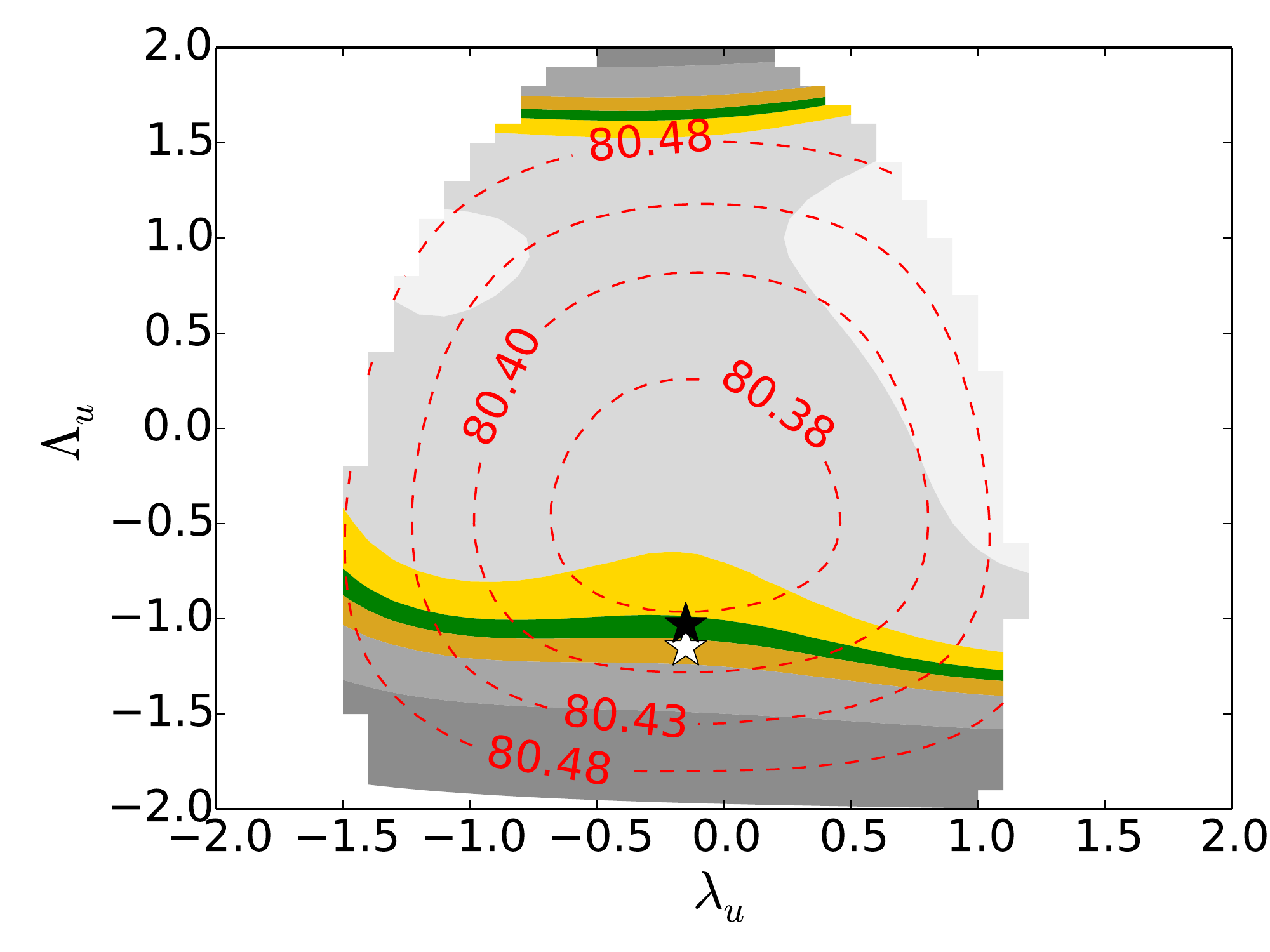}
\end{minipage}
\begin{minipage}{0.05\textwidth}
\includegraphics[width=\textwidth]{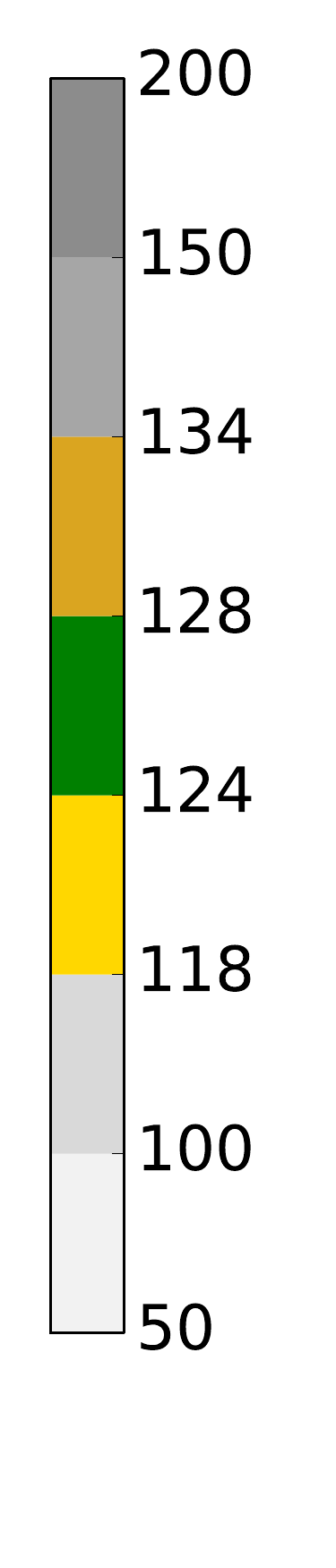}\\
\includegraphics[width=\textwidth]{img/mhmw_colorbar.pdf}
\end{minipage}
\caption{Contour plots showing the behavior of $m_{H_1}$ given by the color map and $m_W$ given by
the red contour lines. 
The plots are ordered horizontally by benchmark points (top row for BMP1, bottom row for BMP3),
while vertically different combinations of model parameters are varied.
The white stars mark the original benchmark points from Ref.~\cite{Diessner:2014ksa}, whereas
the black ones show the adapted points after  taking into account the two-loop corrections.}
\label{img:mhmw1}
\end{figure}

In this section we present an update of the analysis of
Ref.~\cite{Diessner:2014ksa}, using the more precise evaluation of the
Higgs boson mass. Ref.~\cite{Diessner:2014ksa} studied the mass
predictions of the $W$ and lightest Higgs bosons in the MRSSM and
showed that agreement with experimental data is possible, in spite of
tree-level shifts from violations of custodial symmetry and from
mixing with other Higgs states, respectively.

Table~\ref{tab:BMPold} shows benchmark parameter points defined in that
reference. They exemplify parameter regions in which $m_W$ and
$m_{H_1}$ agree with experiment. They are characterized by large
$|\Lambda|\approx 1$, rather light Dirac higgsinos and gauginos, and
they have $\tan\beta=3,10,40$, respectively.

For all three benchmark points the two-loop correction to $m_{H_1}$ is
around $+5$~GeV. As discussed in the previous sections, the largest
part of this is due to the ${\cal O}(\alpha_t\alpha_s)$
corrections. The MRSSM-specific corrections of ${\cal
O}(\alpha_\Lambda^2)$ are small since the values of $\Lambda_u$,
though large, are still not as large as needed to make these
corrections dominate, see
Figs.~\ref{img:dhlamdlamuTest1},~\ref{img:dhlamdlamuTest2} for two out of three benchmarks. 
The magnitude of the total two-loop correction is consistent with the theory
error estimate given in Ref.~\cite{Diessner:2014ksa}. 

The upward shift of $m_{H_1}$ implies that it is easier to obtain
agreement with the measured value, i.e.~smaller values of
$|\Lambda_u|$ are sufficient. In Tab.~\ref{tab:BMPnew} we provide new,
slightly modified benchmark points, whose definitions
differ only in the values of $\Lambda_u$. The two-loop Higgs boson mass
prediction agrees well with experiment, and the good agreement of
$m_W$ with experiment is unchanged. Likewise, both the old and the new
set of benchmark points pass checks
against \texttt{HiggsBounds}~\cite{Bechtle:2008jh,Bechtle:2011sb, 
%Bechtle:2013gu,
Bechtle:2013wla} and \texttt{HiggsSignals}~\cite{Bechtle:2013xfa,Bechtle:2014ewa}.

In Fig.~\ref{img:mhmw1} we give an update to some of the subfigures
from Figs.~4 and 5 of Ref.~\cite{Diessner:2014ksa}. These show the
predictions of $m_W$ and $m_{H_1}$ as contour lines in several
two-dimensional parameter spaces. The Higgs boson mass is evaluated at the
two-loop level. 
As discussed before, with the exception of the regions of very large
$\Lambda$, there is a general positive contribution to
the lightest Higgs boson mass between 4 and 5~GeV. Accordingly, the
contour lines, in particular the central green region in which the
Higgs boson mass agrees with experiment, shift to slightly lower values of
$\Lambda$. 
Also, the overlap region, where Higgs and $W$
boson masses agree with experiment, is enlarged.

\begin{table}[t]
\begin{center}
\begin{tabular}{l|rrr}
%\toprule
&BMP1&BMP2&BMP3\\
\hline
$\tan\beta$  &  3  &  10  &  40\\
$B_\mu$      &  $500^2$  &  $300^2$  &  $200^2$\\
$\lambda_d$, $\lambda_u$&   $1.0,-0.8$ &  $1.1,-1.1$  &   $0.15,-0.15$\\
$\Lambda_d$, $\Lambda_u$&  $-1.0,-1.2$ &  $-1.0,-1.0$ & $-1.0,-1.15$\\
$M_B^D$&$600$&$1000$&$250$\\
$m_{R_u}^2$&$2000^2$&$1000^2$&$1000^2$\\
\midrule
$\mu_d$, $\mu_u$&\multicolumn{3}{c}{$400,400$}\\
$M_W^D$&\multicolumn{3}{c}{$500$}\\
$M_O^D$&\multicolumn{3}{c}{$1500$}\\
$m_T^2$, $m_S^2$, $m_O^2$&\multicolumn{3}{c}{$3000^2,2000^2,1000^2$}\\
$m_{Q;1,2}^2$, $m_{Q;3}^2$&\multicolumn{3}{c}{$2500^2,1000^2$}\\
$m_{D;1,2}^2$, $m_{D;3}^2$&\multicolumn{3}{c}{$2500^2,1000^2$}\\
$m_{U;1,2}^2$, $m_{U;3}^2$&\multicolumn{3}{c}{$2500^2,1000^2$}\\
$m_L^2$, $m_E^2$&\multicolumn{3}{c}{$1000^2$}\\
$m_{R_d}^2$&\multicolumn{3}{c}{$700^2$}\\
\midrule
$v_S$& $4.96$ & $0.67$ &$-0.30$\\
$v_T$&$-0.34$ &$-0.20$ &$-0.34$\\
$m_{H_d}^2$&$673^2$ &$743^2$ &$1160^2$ \\
$m_{H_u}^2$& $-535^2$ &$-542^2$ & $-541^2$ \\
\midrule
$m_{H_1}$ & 130.3 GeV & 130.3 GeV & 129.8 GeV \\
$m_W$ &80.400 GeV & 80.384 GeV & 80.393 GeV\\
\texttt{HiggsBounds}'s \texttt{obsratio} & $0.67$ & $0.68$ & $0.67$\\
 \texttt{HiggsSignals}'s p-value & 0.03 & 0.03 & 0.03 \\
%\bottomrule
\end{tabular}
\end{center}
\caption{Benchmark points of Ref.~\cite{Diessner:2014ksa}. Dimensionful parameters are given in
GeV or GeV${}^2$, as appropriate. The first two parts define input parameters. 
The third part shows  parameters derived from electroweak symmetry breaking after solving the tadpole equations at two loops. The last part gives the theory predictions for the Higgs boson mass at the two-loop level and further quantities relevant for comparison with experiment.}
\label{tab:BMPold}
\end{table}
\begin{table}[t]
\begin{center}
\begin{tabular}{l|rrr}
%\toprule
&BMP1'&BMP2'&BMP3'\\
\hline
$\Lambda_u$&  $-1.11$ & $-0.85$ & $-1.03$\\
\midrule
$v_S$& $5.2$ & $1.01$ &$-0.22$\\
$v_T$&$-0.25$ &$-0.02$ &$-0.21$\\
$m_{H_d}^2$&$674^2$ &$764^2$ &$1160^2$ \\
$m_{H_u}^2$& $-502^2$ &$-512^2$ & $-516^2$ \\
\midrule
$m_{H_1}$ & 125.3 GeV & 125.5 GeV & 125.4 GeV \\
$m_W$ &80.397 GeV & 80.381 GeV & 80.386 GeV\\
\texttt{HiggsBounds}'s \texttt{obsratio} & $0.61$ & $0.65$ & $0.87$\\
 \texttt{HiggsSignals}'s p-value & 0.72 & 0.66 & 0.72 \\
%\bottomrule
\end{tabular}
\end{center}
\caption{Adapted benchmark points; other parameters are as given in Tab.~\ref{tab:BMPold}}
\label{tab:BMPnew}
\end{table}
%\FloatBarrier

%% file: tex/conclusions.tex
\section{Conclusions}
\label{sec:conclusions}

In this work we have presented the impact of two-loop corrections on
the mass of the lightest Higgs boson in the MRSSM.
The calculation has been performed using the framework of \texttt{SARAH}
in the approximation of the vanishing electroweak gauge couplings and
external momenta of the Higgs self energies.
The code has been cross-checked with an analytic calculation of the
most important new corrections. We have separately analyzed the impact of
contributions involving the $\lambda,\Lambda$-couplings, which already
appear in the one-loop corrections, and of the strong corrections
involving gluon, Dirac gluino, and sgluon exchange.

In the previous work \cite{Diessner:2014ksa} and the present paper we
have found that the lightest Higgs boson mass in the MRSSM differs
from the one in the 
usual MSSM in several respects. At tree-level the additional mixing
with additional scalar states reduces the MRSSM Higgs mass below the
MSSM value. At the one-loop level, the top/stop contributions cannot
be as large as in the MSSM, because stop mixing is forbidden by
R-symmetry. However, the new contributions from the superpotential
$\lambda,\Lambda$-terms have a similar structure as the top/stop
contributions. If the $\lambda,\Lambda$-couplings are similar in
magnitude to the top Yukawa coupling, the lightest Higgs boson mass
can easily be in the ballpark of the experimentally allowed range.

The two-loop corrections governed by these
$\lambda,\Lambda$-couplings, however, amount to only 1~GeV or less in 
parameter regions in which the Higgs boson mass agrees with
experiment. The most important two-loop contributions are the strong
corrections of ${\cal O}(\alpha_t\alpha_s)$. As we have shown the
Dirac gluino and gluon contributions alone are very similar to the
MSSM strong contributions for vanishing stop mixing. The inclusion of
the sgluons changes the picture. The sgluon contributions are positive
and rise with the Dirac gluino mass, such that the total
 ${\cal O}(\alpha_t\alpha_s)$ corrections of the MRSSM are larger than
the ones of the MSSM, independently of the magnitude of stop mixing.

Overall, the MRSSM two-loop corrections to the lightest Higgs boson
mass are typically positive. E.g.\ for the benchmark parameter points
proposed in Ref.~\cite{Diessner:2014ksa}, the two-loop
corrections to the Higgs boson mass amount to approximately
$+5$~GeV, within the error estimate of that reference. Since
perturbation theory shows a converging behavior and 
since the $\lambda,\Lambda$-corrections are subdominant (for
$|\lambda|,|\Lambda|$ less than around 1.2), we estimate the remaining theory
uncertainty to be not larger than the one of the MSSM.
 
The positive two-loop corrections make it easier to achieve agreement
between the theory prediction for the lightest Higgs boson mass and
the measured value. We have provided an update of the analysis of 
Ref.~\cite{Diessner:2014ksa}, showing parameter regions of
simultaneous agreement of the Higgs and W boson mass predictions with
experiment. Compared to Ref.~\cite{Diessner:2014ksa}, the allowed
parameter regions are slightly larger and located at smaller values of
the $\lambda,\Lambda$-couplings.